\pdfoutput=1
\documentclass[twocolumn,floatfix,superscriptaddress,aps,pre]{revtex4-1}

\usepackage{graphicx}% Include figure files
\usepackage{dcolumn}% Align table columns on decimal point
\usepackage{bm}% bold math
\usepackage{amsmath}
\usepackage{natbib}
\usepackage{placeins}

\begin{document}

\preprint{APS}
\raggedbottom
\title{Crowder and Surface Effects on Self-organization of Microtubules}

\author{Sumon Sahu}
\affiliation{Department of Physics, Syracuse University}
\author{Lena Herbst}
\affiliation{Department of Microbiology, UMass Amherst}
\author{Ryan Quinn}
\affiliation{Department of Biochemistry and Molecular Biology, UMass Amherst}
\author{Jennifer L. Ross}
\email{Corresponding author\\Email address: jlross@syr.edu}
\affiliation{Department of Physics, Syracuse University}

\begin{abstract}
Microtubules are an essential physical building block of cellular systems. They are organized using specific crosslinkers, motors, and influencers of nucleation and growth. With the addition of anti-parallel crosslinkers, microtubule self-organization patterns go through a transition from fan-like structures to homogeneous tactoid condensates in vitro. Tactoids are reminiscent of biological mitotic spindles, the cell division machinery. To create these organizations, we previously used polymer crowding agents. Here we study how altering the properties of the crowders, such as type, size, and molecular weight, affect microtubule organization. Comparing simulations with experiments, we observe a scaling law associated with the fan-like patterns in the absence of crosslinkers. Tactoids formed in the presence of crosslinkers show variable length, depending on the crowders. We correlate the subtle differences to  filament contour length changes, affected by nucleation and growth rate changes induced by the polymers in solution. Using quantitative image analysis, we deduce that the tactoids differ from traditional liquid crystal organization, as they are limited in width irrespective of crowders and surfaces, and behave as solid-like condensates.

\end{abstract}
\maketitle

\section{\label{sec:level1} Introduction}
Cytoskeletal systems inside a cell, composed of actin, microtubules, and their associated proteins, are dynamic and complex in nature. The organization and rapid reorganization of cytoskeletal fibers are responsible for various crucial jobs in cells such as cell division, morphogenesis, and cell motility. One essential microtubule-based organization is the mitotic spindle. This dynamic, self-organized machine spontaneously forms around condensed and duplicated chromosomes in order to align and eventually separate the DNA into two daughter cells during cell division. The physics behind spindle formation, steady-state behavior, and transition into later stages of cell division is of primary interest to cell biology, biophysics, and soft condensed matter physics, as it could reveal important information to active matter physics \cite{oriolareview18}.

The spindle-shape and fluidity of the mitotic spindle is reminiscent of the tactoid liquid crystalline phase \cite{Brugues2014PhysicalSelf-organization}. Since the spindle is made from microtubule filaments, which have a high aspect ratio, it is reasonable to apply liquid crystal models and theories. Tactoids are condensed phases of aligned liquid crystal molecules in a background of isotropic molecules. They can be bipolar or homogeneous, depending on the director field within the condensate. Bipolar tactoids have two point defects on the surface (boojums) at opposite ends, while homogeneous tactoids' defects exist at infinity, so that the director vector field is constant within the condensate. Liquid crystal theory has been used previously to explain the shape and director field inside tactoids \cite{Kaznacheev2002TheCrystals,Prinsen2003ShapeTactoids}. The organization is a result of the interplay between the bulk elastic energy and anisotropic surface energy. 

Tactoid organization has been observed in a variety of biological systems composed of elements with high aspect ratio, including actin \cite{Weirich2017LiquidBundles,Oakes2007GrowthF-actin,danielscheff2020}, amyloid fibrils \cite{Nystrom2018Confinement-inducedTactoids,Bagnani2019AmyloidTactoids}, tobacco mosaic viruses (TMV) \cite{bernal1941x, Maeda1997AnEvaluation}, and fd viruses \cite{Dogic2001DevelopmentTransition,Dogic2003SurfaceRods}. Condensation of biological tactoids can be driven by specific crosslinking \cite{Weirich2017LiquidBundles}, macromolecular crowding \cite{Tortora2010Self-assemblyAdditives}, and high density of particles \cite{Onsager1949TheParticles,Oakes2007GrowthF-actin}, which are passive, entropic drivers of self-assembly. 

We have previously shown that microtubules can also condense into tactoids with the addition of a passive microtubule crosslinker, MAP65, a member of the  MAP65/Ase1/PRC1 family used in spindle organization in plants, mammalian, and yeast cells \cite{Edozie2019Self-organizationStructures}. This family of crosslinkers has been found in the mitotic spindle midzone in metaphase, telophase, and cytokinesis \cite{Zhu2006SpatiotemporalCells, Pringle2013MicrotubuleMAP65}. MAP65 binds specifically to anti-parallel microtubule overlaps via a single microtubule-binding domain with a self-associating ``crossbridge region'' that is flexible,  about 25 nm in length \cite{Subramanian2010InsightsProtein,Li2007TheDimer, Chan1999TheMicrotubules}. MAP65 binding is comparatively low affinity $K_{D}\sim1.2~\mu M$ (low dwell time on microtubule) with contact angle dependent binding \cite{Tulin2012Single-moleculeBundling}. Interestingly, it has been shown that MAP65 promotes tubulin assembly, influencing nucleation by reducing the critical concentration of tubulin required for spontaneous filament nucleation \cite{Li2007AtMAP65-1Assembly, Mao2005TwoMicrotubules}. In our prior work, we found that microtubule tactoids were homogeneous, nematic tactoids  \cite{Edozie2019Self-organizationStructures}. Unlike other biological tactoids or the mitotic spindle itself, our homogeneous microtubule tactoids were not able to coalesce or internally rearrange, as would be expected for a liquid \cite{Edozie2019Self-organizationStructures}, leading us to conclude that they were jammed. Another interesting observation was that these microtubule tactoids displayed a limited width. 

Our prior work was performed in specific conditions, leading us to speculate that the jammed nature of the tactoids could be due to the experimental methods we used. Specifically, we used polymers in solution to aid in condensation and polymers on the surface of the experimental chamber to eliminate protein adhesion. We supposed that these large, complex polymers could alter the filament interactions to reduce mobility of the microtubule organizations.

In order to explore these possibilities, we quantify microtubule organization in the presence and absence of MAP65 with altered macromolecular crowding agents and surface treatments. We perform new experiments as well as simulations to understand the mechanism of microtubule pattern formation and scaling associated with these patterns. Overall, we find that the microtubule contour length and the presence of the MAP65 have the largest effects on the organizations we observe. We find that the crowding agents have mild affects on patterns, most likely due to polymer effects on microtubule nucleation and growth, which affect microtubule contour length. The surface interaction is important in the absence of MAP65, when microtubules self-organize onto the surface, driven there by depletion interactions. These new studies demonstrate the reproducibility of microtubule tactoids, which continue to be homogeneous and jammed with fixed width. \\ 

\section{\label{sec:level2} Methods and Techniques}
 
{\bf Tubulin preparation}: Unlabeled and fluorescently labeled (Dylight 488) lyophilized tubulin from porcine brain is purchased from Cytoskeleton. Tubulins are resuspended to 5 mg/ml in PEM-80 buffer (80 mM PIPES, pH 6.8, 1 mM MgSO$_{4}$, 1 mM EGTA). Fluorescent and unlabeled tubulin are combined to a final labeling ratio of $\sim 4\%$ fluorescently labeled tubulin. Tubulin is aliquoted, drop-frozen, and stored in -80 $^{o}$C for later use. Aliquots are thawed on ice prior to use.\\

{\bf MAP65}: The microtubule crosslinker, MAP65-1 (MAP65) plasmid was a gift from Ram Dixit (Washington University, St. Louis). The full protein purification protocol is detailed in \cite{Stanhope2015MicrotubulesPatterns,Tulin2012Single-moleculeBundling}. Briefly, protein is expressed using {\it Escherichia coli} BL21(DE3) cells, grown to an OD$_{600}$ of 1, lysed, and clarified. MAP65 protein is recovered from the lysate through affinity between the 6$\times$ histidine tag and Ni-binding substrate \cite{Stanhope2015MicrotubulesPatterns}. Purified protein is checked on an SDS-PAGE gel, aliquoted, and stored at -80 $^{o}$C.\\   

{\bf Crowding agents}: Stock solutions of macromolecular crowders used for these experiments are dissolved in PEM-80 buffer or dd$H_{2}O$ (water). The polymers we used are: 8 kDa Polyethylene glycol (8 kDa PEG), 14 kDa Methylcellulose (14 kDa MC), 88 kDa Methylcellulose (88 kDa MC), and 100 kDa Polyethylene glycol (100 kDa PEG). Stock solutions of each polymer are as follows: 20$\%$ (w/v) 8 kDa PEG in PEM-80, 2.4$\%$ (w/v) 14 kDa MC in water, 3$\%$ (w/v) 88 kDa MC in water, and 5$\%$ (w/v) 100 kDa PEG in PEM-80. For experiments, outlined below, each of these stock solutions are diluted in the experimental mix to 1$\%$ (w/v) for 8 kDa PEG, 0.12$\%$ (w/v) for 14 kDa MC, 0.15$\%$ (w/v) for 88 kDa MC, and 0.25$\%$ (w/v) for 100 kDa  PEG. 

We selected to use these polymers because the PEG is a commonly-used polymer in biological and soft matter systems as an inert crowder. It is a linear polymer that is highly studied so much is known about how the molecular weight scales with size and viscosity. The MC, which is a large, branched polymer used in food processing and gaining popularity in soft matter experiments as a crowder. Because we previously used this polymer in our last study \cite{Edozie2019Self-organizationStructures}, we wanted to continue to perform experiments with the original polymer as a control to ensure reproducibility.

The concentration of 88kDa MC is set because that was the amount we used previously, and we use this as our control parameter to ensure reproducibility and make comparisons to our prior work. For the new crowding agents, we seek to alter the molecular weight while maintaining the polymer in the dilute regime. Each polymer has a different critical concentration,  $c^*$, that denotes the transition from dilute to semi-dilute regime. The concentrations we chose are to ensure that the polymers stayed in the same dilute regime, below their respective critical concentrations. These parameters of each polymer we used are reported in Table IV in Appendix A. 

The second constraint is to have a lower viscosity due to the polymers in these new experiments. The control 88 kDa MC has a measured viscosity of $\sim 2$~cP, while we attempted to have the other crowders at a viscosity of $\sim 1$~cP. We directly measured the kinematic viscosity of the polymer solutions as well as dd$H_{2}O$ using an Ubbelohde glass capillary viscometer (Cannon Instrument) at 37 $^{o}$C using a hot water bath. Kinematic viscosity (cSt) is multiplied by the density of the solutions (g/ml) to produce dynamic viscosity (cP) (viscosity used in this paper). The dd$H_{2}O$ viscosity we measure at 37 $^{o}$C was 0.77 cP. These viscosities are reported in Table IV in Appendix A.\\

{\bf Silanized coverslips:} Coverslips are treated to inhibit protein binding to the surface using a block co-polymer brush of Pluronic-F127, as previously described \cite{Edozie2019Self-organizationStructures}. Coverslips are cleaned with ethanol, acetone, potassium hydroxide (KOH) from Sigma, and treated with  2$\%$ (w/v) dimethyldichlorosilane solution (GE Healthcare) to make the surface hydrophobic. The full silanization protocol can be found in \cite{dixitross2010}. Just prior to use, these cover slips are coated with the block copolymer, Pluronic F127, to create a polymeric brush.\\

{\bf Lipid surface preparation:} For samples with lipid bilayer surface coatings, we create small unilamellar vesicles (SUVs) to coat the surface. SUVs $<$ 100 nm in diameter are made of phospholipid 1-palmitoyl-2-oleoyl-glycero-3-phosphocholine (POPC) (Avanti) suspended in PEM-80 buffer. First, 40 $\mu$l of 10 mg/ml  POPC in chloroform is well mixed with and additional 70 $\mu$l of chloroform. The lipid-chloroform mixture is dried under $N_{2}$ gas and further desiccated inside a vacuum desiccator for $>$15 min.  Dried lipid is resuspensed with PEM-80 buffer by using a vortex mixer for a minute to form giant unilamellar vesicles (GUV), which are sonicated for three minutes to form SUVs. The SUV solution is kept sealed with parafilm in 4 $^{o}$C for use over one week. For experiments with labeled lipid bilayer surface, we use the same procudre with 1 mg/ml 18:1 Liss Rhod PE (Avanti) in chloroform. \\

{\bf Flow chamber preparation:} Experimental flow chambers with an approximate volume of 10 $\mu$l are made from a cleaned glass slide (Fisher), glass cover slip (Fisher), and double-sided tape (3M). Glass slides are cleaned with 70$\%$ ethanol and water (milliQ) and dried using Kimwipes. 

For samples with polymer surface coatings, silanized cover slips are used to create the chamber. The cover slip is coated with 5$\%$ Pluronic F127 by flowing into the chamber and incubating for 7-10 minutes in a humid environment to avoid evaporation. For samples with lipid bilayer coated surfaces, cover slips and slides are cleaned with Ultraviolet Ozone (UVO) cleaner for 10 minutes prior to flowing in small unilamellar vesicles (SUVs). Flow chambers are incubated in humid chambers to allow SUVs to adhere to the surface for $\sim$ 10 minutes. Chambers are washed with PEM-80 buffer to remove excess SUVs from solution.

After surface treatments are performed, the experimental assay mix (see below) is added, and chambers are sealed immediately afterwards with epoxy to prevent evaporation and flow in the sample. One chamber is used to inspect the experiment using fluorescence imaging (see below) while additional chambers are incubated at 37 $^{o}$C for $1 \sim 3$ hours.\\

{\bf Experimental assay mix:} The experimental assay mix contains tubulin, crowding agents, and crosslinkers as specified for each experiment. Each experimental assay mix contains 13.6 $\mu M$ tubulin fluorescently labeled (4$\%$ labeling ratio), drop frozen, and thawed on ice. To enhance microtubule nucleation and stabilize microtubules, 1 mM GMPcPP (Jena Bioscience) is added. For experiments in polymer coated chambers, 0.5$\%$ F127 is added to maintain polymer surface coating. To limit photobleaching and photodamage during the experiment, 25 mM DTT is added with 0.25 mg/ml glucose oxidase, 0.075 mg/ml catalase from bovine liver, and 7.5 mg/ml glucose (all reagents from Sigma) as an oxygen scavenging system.  Oxygen scavengers are added last, just prior to mixing and flowing into the chamber. 

Specific crowders are added to the experimental mix as specified above.  MAP65 crosslinkers are added to the solution, according to the percentage bound to tubulin: 0$\%$, 3$\%$, and 10$\%$ which corresponds to 0, 0.44 $\mu M$, and 1.49 $\mu M$ final concentration \cite{Edozie2019Self-organizationStructures}.

We perform comparison experiments using the same reagents of proteins and chemicals in order to ensure reproducibility. Specifically, direct comparisons are made on the same week, using the exact same tubulin and MAP65 aliquots. This ensures the most direct comparison for different experiments. \\

{\bf Imaging:} Imaging of microtubule organizations is performed using total internal reflection fluorescence (TIRF) microscopy on an inverted Nikon Ti-E microscope with Perfect Focus. The TIRF is illuminated using a 488 nm diode laser aligned into the epi illumination path \cite{dixitross2010}. Image data is taken using a 60x oil-immersion (1.49 NA) objective, expanded with a 2.5x lens onto an Andor Ixon EM-CCD camera. The image scale is 108 nm per pixel. Images are displayed and recorded using the Nikon Elements software. Time series images are taken every 2 seconds for 1 hour. Still images of chambers are imaged within 1 $\sim$ 3 hrs.

Fluorescence recovery after photobleaching (FRAP) is performed using spinning disc microscopy (Yokogawa W1) on an inverted Nikon Ti-E microscope with Perfect Focus and 100x oil immersion (1.49 NA) objective imaged with a Andor Zyla CMOS camera. Confocal image acquisition and photobleaching were performed with 561 nm laser and 405 nm laser with orthogonal stimulation respectively. The data was taken as follows: 1) Acquisition: 15 seconds with 2 seconds interval, 2) Photobleaching: 5 seconds, 3) Acquisition: 180 seconds with 2 seconds interval.\\

{\bf Turbidity measurements:} Kinetics of tubulin polymerization in presence of crowding agents is quantified by absorbance measurements at $\lambda=350$nm, $A_{350}$. A flat bottom 96-well plate (Fisher) with experimental assay mix is placed in the microplate reader SpectraMax i3 (Molecular Devices) at 37 $^{o}$C temperature to induce polymerization. Data is taken for 50 minutes in 45 seconds intervals. Each experimental configuration is performed in triplicate to check reproducibility of the results. We have six different experimental configurations: 1) negative control sample, to normalize our data, contains PEM-80 buffer, DTT, oxygen scavenger, glucose, and tubulin with final concentration 13.6 $\mu$M but no nucleotide to initiate nucleation and growth of microtubules, 2) positive control has the same reagents as the blank with GMPcPP added to induce microtubule polymerization, 3) experimental tests with same reagents as the positive control including a) 1$\%$ (w/v) for 8 kDa PEG, b) 0.12$\%$ (w/v) for 14 kDa MC, c) 0.15$\%$ (w/v) for 88 kDa MC, and d) 0.25$\%$ (w/v) for 100 kDa  PEG. 

Each curve is normalized to their average final and  initial absorbance values and fit to the Boltzmann sigmoid equation \cite{schummel2017},
\begin{equation}
y= a - \frac{b}{1+\exp\left({\frac{t-t_{0}}{\tau}}\right)} \label{eq:boltzman}
\end{equation}
where a is the scaled final absorbance value (a $\sim 1$ after normalizing), b is the difference between scaled final and initial absorbance value (b $\sim 1$ after normalizing), $t_{0}$ is the inflection point, and $\tau$ is the characteristic time scale the signifies the flatness of the curve. The kinetics of the polymerization process can be captured in two parameters, the lag time for nucleation $t_{lag}=t_{0}-2\tau$ and the $\tau$ itself. \\

{\bf Data Analysis:} Self-organization of microtubules without crosslinkers are studied by orientation domain analysis using a FIJI plugin called OrientationJ and further analyzed by custom MATLAB scripts, similar to prior methods \cite{Edozie2019Self-organizationStructures}. First, raw gray scale images are smoothed two times in FIJI to reduce noise and pixelated regions. Next, we employ OrientationJ to the image to create a colored image map, where the color indicates the directionality of microtubules or microtubule bundles within the image. The settings used to create color maps are a 5$\times$5 Gaussian window with cubic spline gradient scheme. Color saturation and brightness are kept constant. The color range [0, 255] of the resulting image map is equivalent to an angle range of [0, $\pi$]. In order to determine the size of domains pointing at similar angles, we use the color threshold command in FIJI to select regions in the following four color ranges:  [0, 63], [64, 127], [128, 191], [192, 255]. The image map is turned into four individual orientation region maps by binarizing the color thresholded image maps. For each pixel (x,y) in the original image, the location is assigned to one of the four thresholded image maps and assigned a directionality index, m = 1, 2, 3, 4 corresponding to the four different directionality/color bins ([0, 63], [64, 127], [128, 191], [192, 255]). 

The four m-type binary images are analyzed using moment analysis and spatial auto correlation function analysis to determine the area parameters of the domains pointing in each direction. We use MATLAB's built-in analysis tool, region properties (regionprops) that returns the number and size of domains present in a binary image. We use definitions from percolation theory to describe our domains or clusters \cite{stauffer1992introduction}. We define the domain size distribution, as: 
\begin{equation}\label{eqn1}
n_s =  \frac{{\rm \Lambda}}{L^{2}}   
\end{equation}
where L is the size of our system or window size, and $\Lambda$ is the number of domains with area, s, within a window of size, $L^2$. The values of L for experiment and Cytosim simulations are 55.3 $\mu m$ and 50 $\mu m$ respectively, so this rescaling is used for comparisons. The average of $n_{s}$ over all binary images produce $\overline{n}_{s}$, which is used to compare between different data sets. 

The $0^{th}$ moment of the $n_{s}$ distribution is the number of m-type domains in an binary image, defined as:
\begin{equation}
N = \sum_{s=s_{min}}^{s_{max}} n_{s}    
\end{equation}
The $1^{st}$ moment of the $n_{s}$ distribution defines the probability of m-type domains appearing in an image as: 
\begin{equation}
p = \sum_{s=s_{min}}^{s_{max}}s n_{s}    
\end{equation}  
Since we quantize our continuous rotation symmetry to four values, we should expect $p\sim0.25$. 

The average domain size $\langle s\rangle$ in an image of window size $L^2$ is defined by 
\begin{eqnarray}
\langle s\rangle &=& \frac{\sum_{s=s_{min}}^{s_{max}}s^{2} n_{s}}{p}
\end{eqnarray}
Here values of $s_{min}$ and $s_{max}$ are 1 $\mu m^{2}$ and 2500 $\mu m^{2}$, respectively. The value of $s_{min}$ is set by the binning size, which is 1 $\mu m^{2}$.  
  
We can also compute the correlation length for the domain images using the binarized orientation images. The correlation length is found using the radially averaged pair auto-correlation function. Spatial auto-correlations can be calculated using the Fast Fourier Transform (FFT) \cite{Robertson2012TheorySpectroscopy} according to Weiner-Khinchin theorem: 
\begin{equation}\label{eqn2}
G({\bf r})=\frac{F^{-1}(\lvert F(I)\rvert^{2})}{\rho^{2}M({\bf r})} 
\end{equation}
where $F(I)$ is the FFT of a binarized image, $I$, $\lvert F(I)\rvert^{2}$ is the power spectrum of the FFT of $I$, and $F^{-1}(I)$ is the inverse FFT of $I$. To avoid artifacts due to the periodic nature of the FFT, the images $I$ are increased in size around the boundary by adding 200 additional pixels (experiment) or 300 pixels (cytosim) that are all set to zero (black) \cite{Veatch2012CorrelationOver-counting}. These additional number of pixels are chosen to be consistent with the rescaling.  

In the denominator, $\rho$ is the intensity per unit area in the image. Since the image is binarized, $\rho$ exists between 0 and 1.  The form of $M({\bf r}) = F^{-1}(\lvert F(W)\rvert^{2})$, where W is the window function used for normalization as described in \cite{Veatch2012CorrelationOver-counting}. Specifically, $W$ is an image of the same size as I (with padding) where all pixels are set to white (intensity equals 1).  $G({\bf r})$ is averaged in the azimuthal direction to give $\tilde{g}(r)=1+Ag(r)$, where A is the amplitude. For each binary image, $g(r)$ is fit with a sum of two exponentials $y(r)=ae^{-br}+ce^{-dr}$. In order to find a single correlation length, $\xi$, we determine the value of the radius where $y(\xi)=0.37$, using the best fit equations. For each experimental parameter, $g(r)$ and $\xi$ are averaged over several independent images to give $\overline{g}(r)$ and $\overline{\xi}$ respectively which are used to compare between data sets.

Tactoid growth dynamics are analyzed using the MATLAB function regionprops on cropped binary images created using the Otsu global thresholding algorithm \cite{otsu1979}. The area and major and minor axis lengths are determined for each frame and plotted over time. Intensity profiles along the major axis length are plotted to visualize the change over time. Each video is one hour long and data was averaged over three tactoids.

Microtubule tactoids made in the presence of $10\%$ MAP65 are imaged as still frames and analyzed in FIJI by hand to find the length and width. For comparison, we also use homemade MATLAB scripts using binarized images and the regionprops functions.\\
%MATLAB tactoid measurements appears to be wider in width and shorter in length than hand measurement as there is ambiguous intensity variation along the tactoid boundary, especially at the tip. 

{\bf Cytosim:}
Agent based modeling is performed using an open source cytoskeletal filament simulating package, Cytosim \cite{Nedelec2007CollectiveFibers}. Parameters used for the simulation are given in TABLE \ref{tab:table2}.  We choose the total filament length to represent the experimental concentration, which is controlled by the packing fraction $\phi$ \cite{Rickman2019EffectsSelf-organization}. The packing fraction is,
\begin{equation}
\phi~=~\frac{2R_{s}}{L^2}l_{total}    
\end{equation}
where L is the window size in the simulation, $R_{s}$ is the steric radius of the fiber, and $l_{total}$ is the final total filament length. Filaments grow until the final total filament length, $l_{total}$ is exhausted and they reach a prescribed final mean filament length, $l_{c}$. Thus, the final total contour length, $l_{total}=Nl_{c}$ is defined by the final mean filament length, $l_{c}$ and the number of initial microtubule seeds, $N$. Filaments are allowed to grow from an initial length of 0.02 $\mu m$ with an initial growing speed of 30 $nm/s$ \cite{Roostalu2018DeterminantsMotors}.  

Because we are inspecting different polymers in the model, and these polymers create a depletion force, we wanted to implement a depletion force into the Cytosim modeling. The best method to do this would be to use a Derjaguin like approximation for Asakura–Oosawa depletion forces. Unfortunately, Cytosim does not have this interaction type available. In order to mimic an effective depletion force, two prior papers \cite{strubing2020,letort2015geometrical} in the literature have used a piece-wise function of repulsive and attractive springs with force, $F(r)=kr$, where r is the distance from the center of the filament. The spring constant depended on $r$ as: 
\begin{equation}\label{eq:depletion}
    k =
\begin{cases}
    k_{s},& \text{if } 0\leq r \leq R_{s}\\
    - k_{d},& \text{if } R_{s}\leq r \leq R_{s}+R_{d}\\
    0,              & \text{otherwise}
\end{cases}
\end{equation} 
where $k_{s}$ is purely repulsive within the steric radius, $R_{s}$, and $k_{d}$ is attractive between the range $R_{s}$ and added depletion radius $R_{d}$ to it.

Because this approximation is non-physical, depletion forces do not increase with increasing distances, like the spring does, we sought to check if this depletion interaction was necessary. We repeated the simulations without the attractive region of the piece-wise function, only the repulsive interaction. When we did this, we found only small effects on the quantitative numbers and no effects on the trends we observed. This is elaborated in Fig. \ref{fig:dep_no_dep}, Appendix \ref{appendixAA}.  

Simulations are run long enough for the filaments to reach their prescribed final total filament length. Simulated filaments did not perform dynamic instability because we set the catastrophe rate to zero. This is done because experiments used GMPcPP, the slowly hydrolyzable analog of GTP, that virtually eliminates the catastrophe rate \cite{Hyman1992RoleGMPCPP}. Since dynamic instability is blocked in our simulation, filaments only grow and the growth curve flattens as they reach steady state using this relationship:
\begin{equation}
v_i~=v_0 \left(1-\frac{l_{total,i}}{l_{total}}\right)    
\end{equation}

where $v_i$ is the instantaneous growth speed, $v_0$ is the initial growth speed, listed in TABLE \ref{tab:table2}, $l_{total,i}$ is the instantaneous total filament length in the simulation, and $l_{total}$ is the final total filament length of the simulation, as described above. All simulations are run for 1600 sec to allow the system to reach steady state. The last frame is rendered as an image for analysis identical to that performed for experimental images.

\begin{table}[!ht]
\caption{\label{tab:table2}
Cytosim parameters
}
\centering
\begin{ruledtabular}
\begin{tabular}{c c}
\multicolumn{1}{c}
\textrm{Simulation parameters}&\textrm{Values}\\
\colrule
Time step & 0.1 s \\
Simulation time & 1600 s\\
Viscosity & 0.0025-2.5 Ns/$m^{2}$\footnotemark[1]\footnotemark[3]\footnotemark[5] \\
Geometry (Periodic) & 50$\times$ 50~$\mu m^{2}$~(2D) \\
\colrule
Packing fraction, $\phi$ & 0.5, 0.75, 1\\
Rigidity & 20 pN$\mu m^2$\footnotemark[1]\\
Initial growth speed, $v_{0}$ &  0.03 $\mu m/s$\footnotemark[3]\footnotemark[4]\\
Catastrophe rate & 0 /s \footnotemark[3]\footnotemark[5]\\
Growing Force & 1.7 pN \footnotemark[3]\\
Final total filament length, $l_{total}$ & 12500, 18750, 25000 $\mu m$\\
Final mean filament length, $l_{c}$ & 2, 4, 6, 8 $\mu m$\\
Filament segmentation & 1$\mu m$ \footnotemark[1]\footnotemark[2]\footnotemark[3]\\
Initial filament length & 0.02 $\mu m$\footnotemark[2]\footnotemark[3]\\
Steric radius, $R_s$ & 0.05 $\mu m$\footnotemark[1]\footnotemark[2]\footnotemark[3]\\
Steric force strength, $k_s$ & 50 $pN/\mu m$\footnotemark[3]\\
Depletion radius, $R_d$ & 0.03 $\mu m$\footnotemark[5]\\
Depletion force strength, $k_d$ & 50 $pN/\mu m$\footnotemark[1]
\end{tabular}
\end{ruledtabular}
\footnotetext[1]{ref. 
\cite{strubing2020}}
\footnotetext[2]{ref. \cite{Rickman2019EffectsSelf-organization}}
\footnotetext[3]{ref. \cite{Roostalu2018DeterminantsMotors}}
\footnotetext[4]{ref. \cite{Rickman3427}}
\footnotetext[5]{This study}

\end{table}

\begin{figure}[!ht]
\includegraphics[scale=0.32]{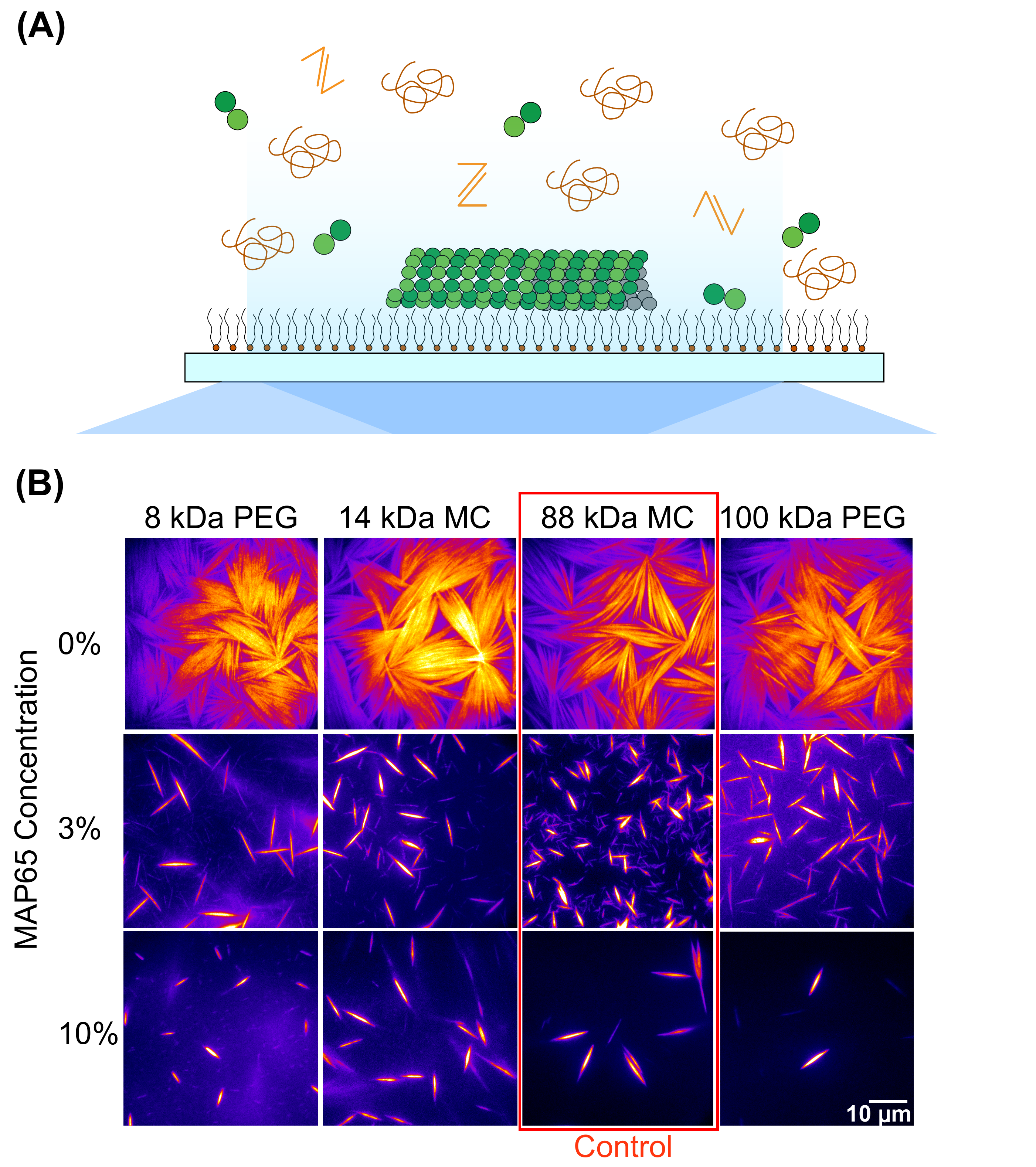}
\caption{\label{fig:fig1} (A) Schematic of experimental system: Tubulin dimers (green) polymerize into microtubles in the presence of crowders (brown polymer balls) and crosslinkers (orange Z) on polymer-coated surfaces imaged with TIRF illumination. (B) Images showing microtubule patterns on polymer brush coated surfaces in the presence of crowders with different molecular weights (8 kDa PEG, 14 kDa MC, 88 kDa MC, 100 kDa PEG) with tubulin concentration 13.6 $\mu$M and MAP65 at 0$\%$ (top), 3$\%$ (middle), 10$\%$ (bottom). Fan-like patterns were observed without MAP65 (top row). High MAP65, 10$\%$, induced tactoid like condensates (bottom row). The red box indicates 88 kDa MC, the condition used in our prior work, and serves as our reproducibility control state \cite{Edozie2019Self-organizationStructures}. Scale bar is 10 $\mu$m for all images.} 
\end{figure}

\section{\label{sec:level3} Results and Discussion}

We seek to understand how the self-organization of microtubules depends on non-specific forces from depletion agents and specific interactions via diffusible crosslinkers. Both mechanisms are responsible for minimizing the microtubule overlap area by contributing to the lateral interaction between them. As we demonstrated previously \cite{Edozie2019Self-organizationStructures}, nucleating and polymerizing microtubules resulted in microtubule self-organizations that showed a transition from fan-like domain patterns to elongated homogeneous tactoids as a function of MAP65 crosslinker concentration. These prior experiments used methylcellulose polymer as a crowder and a polymer-coated surface, which could have affected the organized patterns we observed. The tactoid phase was different from other biological tactoids in two ways: the width of the tactoids was constant and the microtubules inside the tactoids were jammed. 

In order to determine the mechanism driving the pattern formation we previously observed, we have altered the crowding agent type, size, and molecular weight. Qualitatively, for polymer-coated surfaces, all crowders show a similar phase transition, from fan-like structures to tactoids (Fig. \ref{fig:fig1}).

\begin{figure}[!htp]
\includegraphics[scale=0.35]{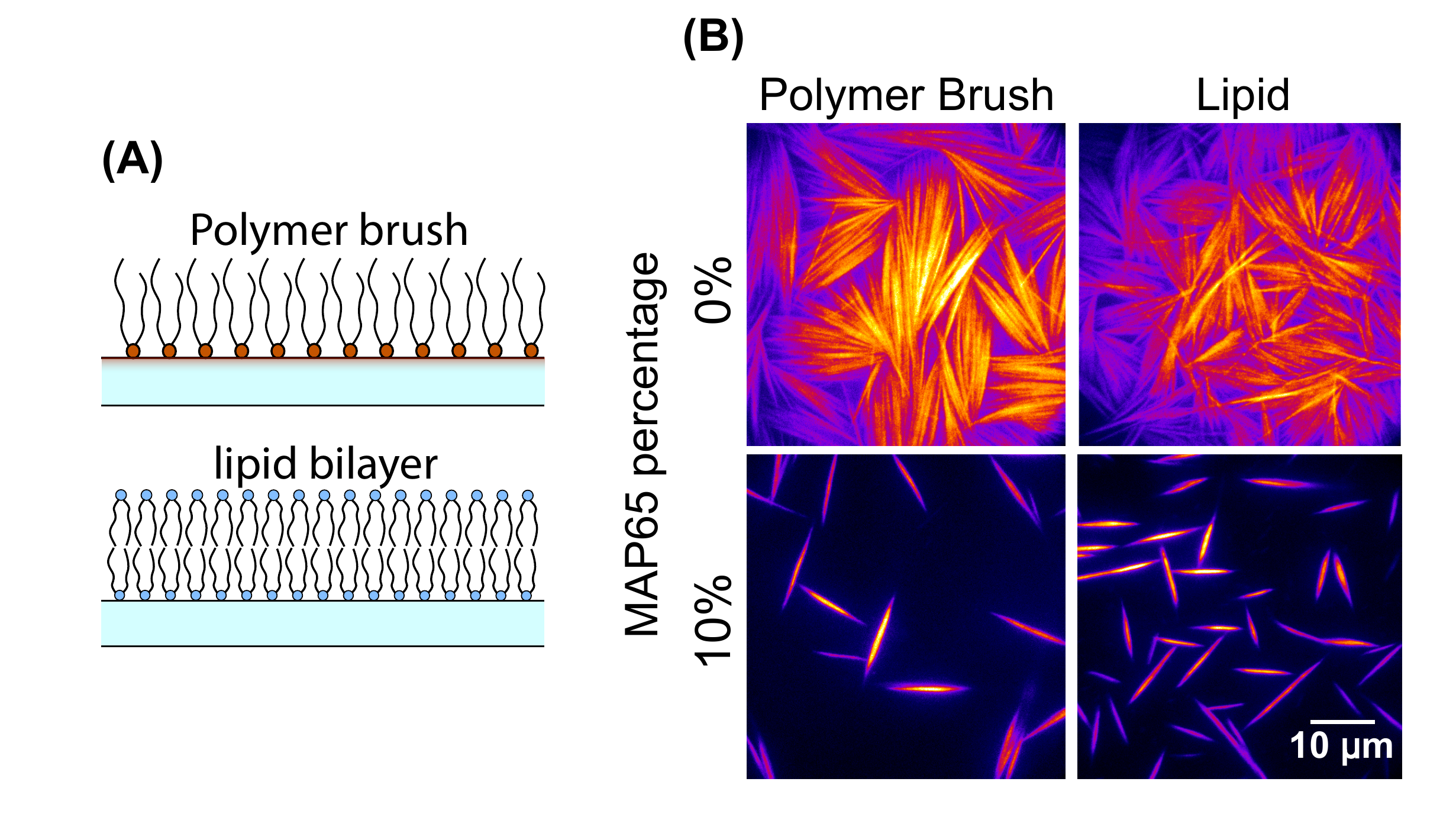}
\caption{\label{fig:fig4-1} Surface effects on patterns. (A) Schematic diagram of F127 polymer brush surface (top) and lipid bilayer surface (bottom). (B) Images of microtubule organizations on polymer-brush surfaces (left column) and lipid surfaces (right column) in the presence of $0 \%$ MAP65 (top row) and  $10 \%$ MAP65 (bottom row). Scale bar is 10 $\mu$m for all images.}
\end{figure}

To test the effect of the surface on microtubule patterns, we repeat the experiments for 0$\%$ and 10$\%$ MAP65 in the presence of 88 kDa MC on a lipid-coated surface. The lipid surface is a fluid bilayer with a slip condition. Qualitatively, we observe that the lipid surface retains the fanlike pattern with increased disorder in the absence of MAP65. Tactoids form robustly on lipid surfaces (Fig. \ref{fig:fig4-1}). 

In an attempt to understand the mechanisms driving the patterns we observe, we simulate the microtubule growth and self-organization using the Cytosim software package (Fig. \ref{fig:fig9}). Simulations allow us to alter the viscosity, depletion interactions, filament length, and growth rates independently and examine the dynamics to steady-state. 

\subsection{Microtubule organization without crosslinkers}

{\bf Polymer surface with different crowders:} Qualitatively, fan-like surface patterns in the absence of crosslinkers are similar for different crowding agents (Fig. \ref{fig:fig1}). To quantify the patterns of microtubule self-organization, we perform a domain analysis similar to our prior publication \cite{Edozie2019Self-organizationStructures}, where the direction of locally-oriented microtubule bundles in the pattern were identified automatically using the OrientationJ plugin in FIJI/ImageJ (Fig. \ref{fig:fig2}A(i)). The image is colored according to the local orientation  (Fig. \ref{fig:fig2}A(ii)) and that color map is used to identify domains where the angles are similar(Fig. \ref{fig:fig2}A(iii)). We threshold the colors to a specific range of colors divided into four different bins that we identify with an index, m (Fig. \ref{fig:fig2}A(iv)). The color/angle thresholding is used to create a binary image where the selected angle range is portrayed in white, and all other angle orientations are in black (Fig. \ref{fig:fig2}A(v)). For each original image of microtubule patterns, we create four binary images highlighting one of the four angle ranges (m = 1,2,3,4). We use the binary images to quantify the domain parameters of the images. Domain parameters of interest included: average domain size (s) distributions $\overline{n}_{s}$ (Fig. \ref{fig:fig2}B) used to determine the average domain size $\overline{\langle s \rangle}$ and the radially averaged spatial mean autocorrelation function, $\overline{g}(r)$ (Fig. \ref{fig:fig2}C) that allows us to determine the mean correlation length $\overline{\xi}$ of the images.

\begin{figure*}[!ht]
\includegraphics[scale=0.5]{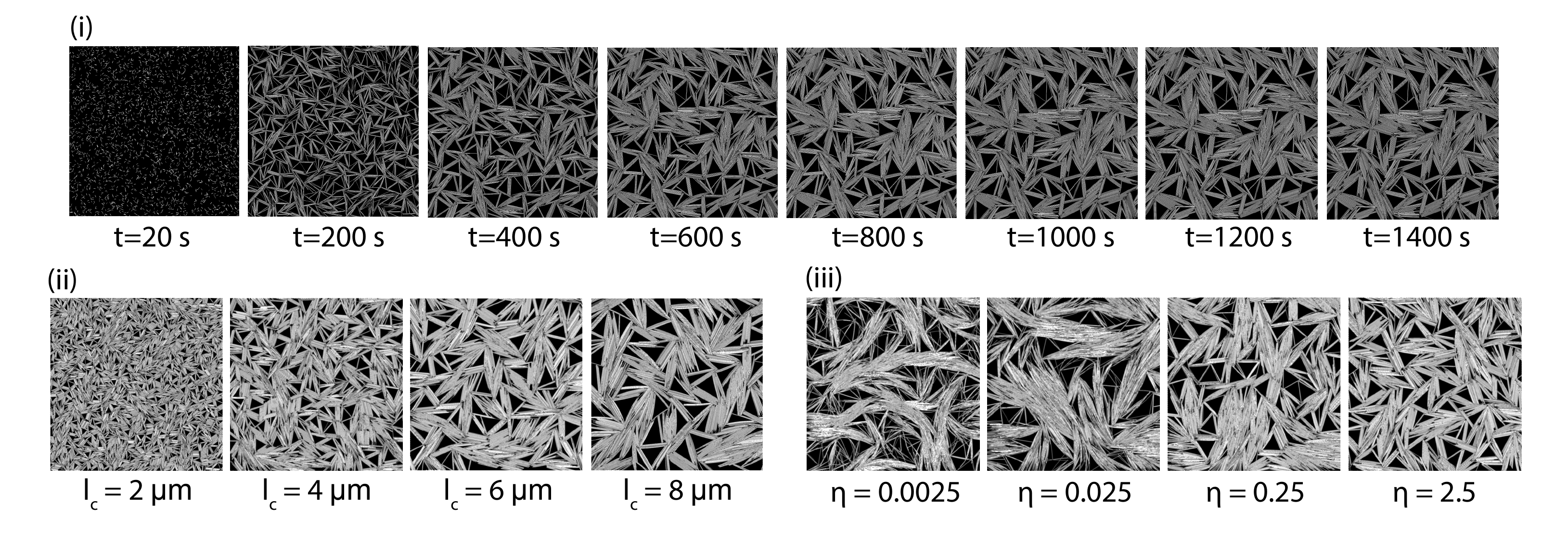}
\caption{\label{fig:fig9} Cytosim simulations of microtubule organization without crosslinkers. (i) Time series of fan-like microtubule pattern growth for parameters, $\phi=0.75$, $\eta=2.5$ Pa-s, $l_{c}=6~\mu m$ with steric interaction and depletion forces present. Filaments grow from initial small seeds until they exhaust the total filament length. (ii) The effect of $l_{c}$ on patterns from left to right: $l_{c}=2,4,6,8~\mu m$ with viscosity $\eta=2.5$ Pa-s and $\phi=0.75$. (iii) The effect of viscosity on steady state patterns from left to right: $\eta=0.0025,~0.025,~0.25,~2.5$ Pa-s with $\phi=0.75$, $l_{c}=6~\mu m$. Final state patterns obtained at higher viscosities match experimental patterns qualitatively.}  
\end{figure*}

Similar to our prior study, we first perform microtubule self-organization experiments using polymer-coated surfaces with four different crowding agents.  We find that the average distributions of the domain areas (divided by the window size, $L^2$ in order to compare with simulations) are all power-laws (Fig. \ref{fig:fig2}B(i)). We test the distributions using the Kolmogorov-Smirnov (KS-test) statistical test and found that distribution for 8 kDa MC is significantly distinct from the rest of the distributions, rejecting the null hypothesis at the 5$\%$ significance level. We find 14 kDa MC and 100 kDa PEG data to be statistically similar (Appendix \ref{appendixB}, TABLE \ref{tab:table5}). When we compare the average domain sizes for these images, the averages are different from one another, although they have large error bars due to `critical-domain' like behavior that results in the power-law (TABLE \ref{tab:table4}).
 
In our prior work, we measured the area distributions for one crowder type, 88 kDa MC, and changed the contour length of the filaments by altering the tubulin concentration. Microtubule contour length was varied by changing the tubulin concentration, $[TUB]$, as described \cite{Edozie2019Self-organizationStructures}. For, $[TUB]=13.6~\mu M,~54.5 ~\mu M,~70 ~\mu M$ median contour lengths were $6.25 ~\mu m,~2.5 ~\mu m, 1.1 ~\mu m$, respectively. The more detailed analysis we are doing here was not performed for the prior experimental results. In order to compare our new experimental and simulation results to our prior results, we re-analyze our prior data (Fig. \ref{fig:fig2}B(ii))\cite{Edozie2019Self-organizationStructures}. The KS-tests show that the area distributions are distinct from one another, rejecting the null hypothesis at the 5$\%$ significance level (Appendix \ref{appendixB}, TABLE \ref{tab:table6}). As reported previously, the average domain size decreases with increasing tubulin concentration (decreasing contour length) (TABLE \ref{tab:table4}) \cite{Edozie2019Self-organizationStructures}.  

Our previous experimental data suggest that the contour length of the microtubule filaments is the control parameter for the domain sizes we observe for microtubule self-organization on polymer coated surfaces without crosslinkers. Unfortunately, in the experiments, we had to increase the tubulin concentration in order to decrease the contour length, which also resulted in a higher number of filaments. Thus, the experiments lacked separate control over the contour length and the number of filaments. In order to directly test the mechanism that contour length is the control parameter for our prior results, here we use Cytosim to simulate polymerizing microtubules with a fixed total number of filaments and final microtubule contour length using the parameters described in the methods and described by prior published works (TABLE \ref{tab:table2}) \cite{Rickman2019EffectsSelf-organization, Roostalu2018DeterminantsMotors, strubing2020}. We plot the domain size distribution as shown in Fig.\ \ref{fig:fig2}B(iii). Examining the average domain size, we recapitulate the dependence on contour length, such that longer contour lengths result in larger average domain sizes (TABLE \ref{tab:table4}). The KS test results show that contour lengths of 2 $\mu m$ and 4 $\mu m$ are distinct from all other distributions, but 6 $\mu m$ and 8 $\mu m$ data are statistically similar, using a 5$\%$ significance level to reject null hypothesis (Appendix \ref{appendixB}, TABLE \ref{tab:table7}). Thus, our simulation agrees with our initial interpretation that the contour length is the likely control parameter for the domain area for orientational domains self-organized on the polymer-coated surfaces. 

We find that the experimental and several of the simulated domain area distributions display a power-law dependence. In critical phenomena theory, power law behavior is achieved when a system approaches a critical point. Our data indicate there is a regime shift for the smallest contour length in simulation data, which did not display a power law. In this case, the distribution starts with a power-law, then falls off exponentially indicating a cut-off cluster size is present in the system. Conversely, configurations which contained large domain areas retain the power law behavior, a signature of critical phenomena. In literature it has been showed that the critical behavior of larger domains result from the finite size of the viewing window \cite{stauffer1992introduction}, and this is likely a factor here.

In order to further characterize the self-organization pattern for microtubules in the absence of crosslinkers, we characterize a different intrinsic length parameter, the correlation length, $\xi$. Using the same binarized images for each orientation domain, we determine the spatial autocorrelation function. Averaging over the different color images, we obtain $\overline{g}(r)$, and plot as a function of the radius from the center of the image (Fig. \ref{fig:fig2}C(i)).  This correlation length, $\xi$, is determined from each autocorrelation function, as described in the methods. These values are averaged over all binary images to define the mean correlation length $\overline{\xi}$ for each experimental parameter (Fig. \ref{fig:fig2}C(i), inset). When we perform the KS test on the distribution of the correlation lengths, $\xi$, we find that the data is not statistically distinct for 8 kDa PEG, 14 kDa MC, and 100 kDa PEG, but the 88 kDa MC $\xi$ values are statistically different from the rest (Appendix \ref{appendixB}, TABLE \ref{tab:table8}).   

In our previous study, we did not perform the correlation length analysis \cite{Edozie2019Self-organizationStructures}, so we seek to determine if this technique agrees with the domain size analysis. As with the domain analysis (Fig.\ \ref{fig:fig2}B(ii)), the auto correlation function (Fig.\ \ref{fig:fig2}C(ii)) and correlation lengths (Fig.\ \ref{fig:fig2}C(ii), inset) appear to depend on the tubulin concentration (TABLE \ref{tab:table4}). Indeed, the KS test results from correlation length distributions show that all the data are statistically distinct from one another (Appendix \ref{appendixB}, TABLE \ref{tab:table9}). 

We perform the correlation length analysis on the simulated microtubule organizations. We find that the auto-correlation profiles, $\overline{g}(r)$ and the mean correlation length $\overline{\xi}$, depend on the filament contour length (TABLE \ref{tab:table4}) and (Fig. \ref{fig:fig2}C(iii)). The simulation shows the same trend observed in experiments, and the KS test results on correlation length distributions show that they are all statistically distinct from one another (Appendix \ref{appendixB}, TABLE \ref{tab:table10}) (Fig. \ref{fig:fig2}C(iii), inset). As above, these results provide added evidence that the contour length is likely the control parameter for the size of the patterns observed in the absence of the crosslinkers. 

\begin{table}[!ht]
\caption{\label{tab:table4}
Fan-like pattern data characteristics
}
\centering
\begin{ruledtabular}
\begin{tabular}{c c c c}
\multicolumn{1}{c}
\textrm{} & \textrm{Configuration}&\textrm{$\overline{\langle s\rangle }$}($\mu m^{2}$)&\textrm{$\overline{\xi}$}($\mu  m$)\\
\textrm{} & \textrm{} & \textrm{mean $\pm$ SD} &\textrm{mean $\pm$ SD}\\
\colrule
 & 8 kDa PEG & $217.8 \pm 144.7$ & $4.2\pm 1.4$\\
$[TUB]$ & 14 kDa MC & $245.7\pm 163.9$ & $4.3\pm 1.5$\\
$=13.6~\mu M$ & 88 kDa MC & $294.4\pm 195.0$ & $5.0\pm 2.0$ \\
& 100 kDa PEG & $208.4\pm 129.9$ & $3.9\pm 0.9$ \\
\colrule
& 13.6 $\mu M$ & $294.4\pm 195.0$ & $5.0\pm 2.0$ \\
88 kDa & 54.5 $\mu M$ & $134.0\pm 72.7$ & $2.7\pm 0.3$ \\
MC & 70 $\mu M$ & $93.3\pm 51.8$ & $2.1\pm 0.5$ \\
\colrule
& 2 $\mu m$ & $37.3\pm 12.5$ & $1.2\pm 0.1$ \\
$\phi=0.75$ & 4 $\mu m$ & $60.5\pm 23.2$ & $1.7\pm 0.2$ \\
& 6 $\mu m$ & $108.3\pm 74.3$ & $2.4\pm 0.3$ \\
& 8 $\mu m$ & $132.4\pm 65.8$ & $2.8\pm 0.5$ \\

\end{tabular}
\end{ruledtabular}
\end{table}

Both the domain area analysis and the auto correlation analysis demonstrate that the contour length of the filaments is the most likely order parameter for the patterns we observe. The evidence for this conclusion is stronger when using the auto correlation lengths, as the errorbars are comparatively smaller. As described above, the power law distribution for the domain areas likely emanates from the cut-off size due to the limited imaging and simulation areas that restrict the domain areas we can measure. For future studies, imaging and simulating over larger areas will allow more accurate measurement of large domains.  

We can quantify the first three moments of $n_{s}$ for the new experimental data, our prior experimental data \cite{Edozie2019Self-organizationStructures}, and new simulated data to compare them. The zeroth moment is the average number of m-type domains $\overline{N}$, the first moment is the mean probability that a position is in one of the four m-type domains $\overline{p}$, and the second moment is the average domain size $\overline{\langle s \rangle}$. We plot each as a function of mean correlation length, $\overline{\xi}$, our independent measure for the images (Fig. \ref{fig:fig11}).

\begin{figure*}[!ht]
\includegraphics[scale=0.45]{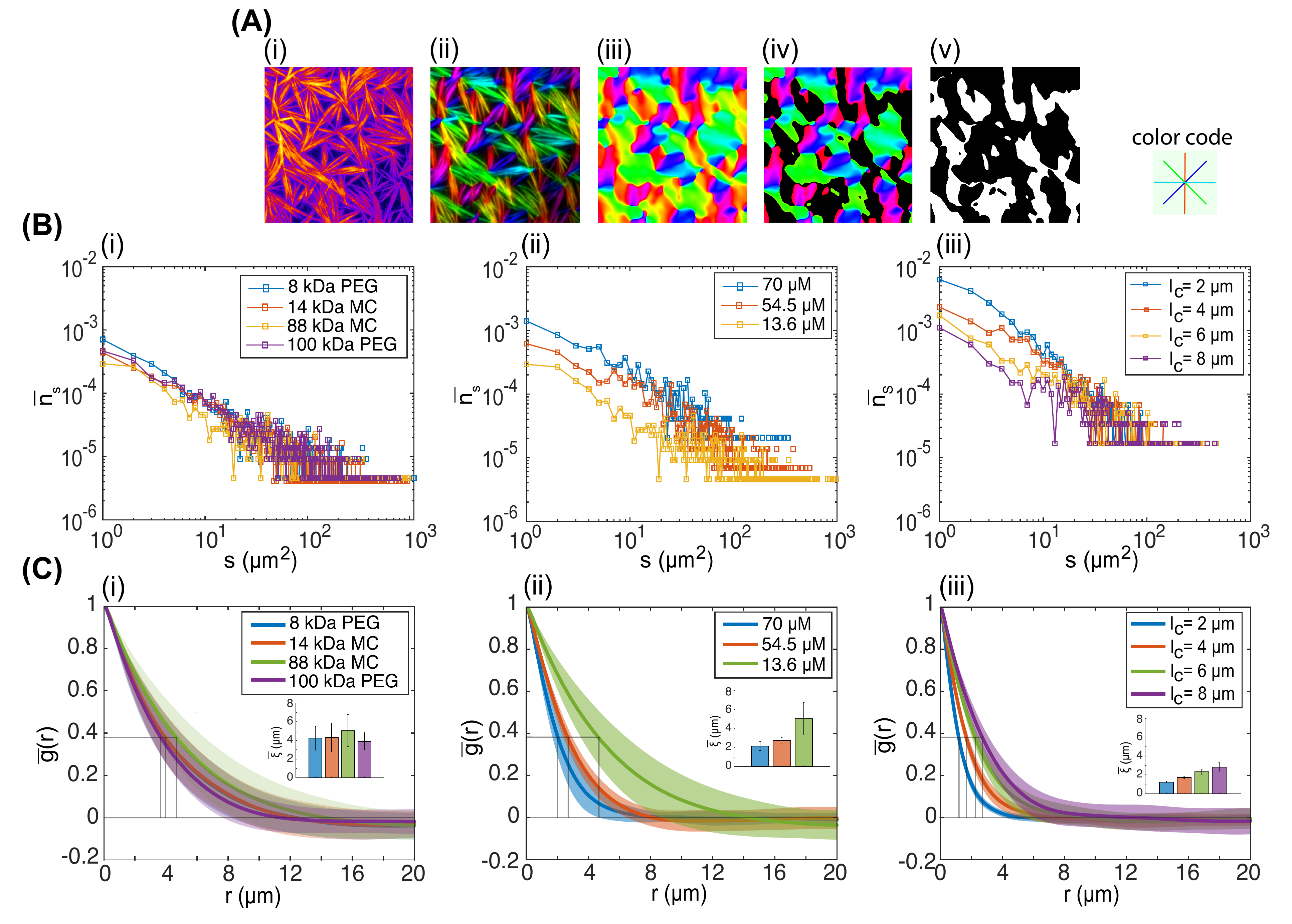}
\caption{\label{fig:fig2}Orientation Analysis: (A)(i) Raw gray scale image with fire look up table. (ii) Image color coded according to orientation of fibers shown. (iii) Color coded regions. (iv) Selected orientation domain shown in black. (v) Binary image created with all regions of same color in white and other regions in black. 
(B) Average domain size distribution $\overline{n}_{s}$.
(i) Different crowders with 13.6 $\mu$M tubulin (n$\geq$ 18) for 8 kDa PEG (blue squares), 14 kDa MC (red squares), 88 kD MC (yellow squares), and 100 kDa PEG (purple squares). 
(ii) Different concentrations of tubulin with MC 88 kDa crowder (n$\geq$ 4) for 70 $\mu$M (blue squares), 54.5 $\mu$M (red squares), and 13.6 $\mu$M (yellow squares). Raw data from \cite{Edozie2019Self-organizationStructures}. 
(iii) Cytosim simulation with four different contour lengths for $\phi = 0.75$ (n=6) $l_c$ is 2 $\mu$m (blue squares), 4 $\mu$m (red squares), 6 $\mu$m (yellow squares), and 8 $\mu$m (purple squares). 
(C) Radially averaged mean auto correlation function, $g(r)$ plots. Lines represent average over all data, $\overline{g}(r)$, and shaded region represents standard deviation of that average. 
(i) Different crowders with 13.6 $\mu$M tubulin (n$\geq$ 18) for 8 kDa PEG (blue line), 14 kDa MC (red lines), 88 kD MC (green line), and 100 kDa PEG (purple lines). Inset data shows the mean correlation length $\overline{\xi}$ with standard deviation (error bars) for 8 kDa PEG (blue bar), 14 kDa MC (red bar), 88 kD MC (green bar), and 100 kDa PEG (purple bar). 
(ii) Different concentrations of tubulin with MC 88 kDa crowder (n$\geq$ 4) for 70 $\mu$M  (blue line), 54.5 $\mu$M  (red line), and 13.6 $\mu$M  (green line). Original raw data from \cite{Edozie2019Self-organizationStructures}. Inset data shows the mean correlation length $\overline{\xi}$ with standard deviation (error bars) for 70 $\mu$M  (blue bar), 54.5 $\mu$M  (red bar), and 13.6 $\mu$M (green bar).
(iii) Cytosim simulation with four different contour lengths length for $\phi = 0.75$ (n=6) $l_c$ is 2 $\mu$m (blue line), 4 $\mu$m (red line), 6 $\mu$m (yellow line), and 8 $\mu$m (purple line).  Inset data shows the mean correlation length $\overline{\xi}$ with standard deviation (error bars) for $l_c$ is 2 $\mu$m (blue bar), 4 $\mu$m (red bar), 6 $\mu$m (yellow bar), and 8 $\mu$m (purple bar).}  
\end{figure*}

The number of domains (zeroth moment) shows a power law dependence on the mean correlation length with an exponent $-1.6\pm 0.1$ (Fig. \ref{fig:fig11}(i), Appendix \ref{appendixB}, TABLE \ref{tab:table17}). It is interesting to note that this dependence is preserved for the simulated data as well. For each data set, the expected contour length of the filaments is also displayed using a color-coded map. The contour-length color-coding shows that the filament length is correlated, with longer contour lengths more likely in the longer correlation lengths and fewer, larger domains. The dependence is not as strongly correlated as the measured mean correlation length $\overline{\xi}$. 

The probability of an m-type domain appearing in an image (first moment) is constant, $\overline{p}\sim 0.25$, no matter that contour length or correlation length (Fig. \ref{fig:fig11}(ii)). As described in the methods, this is expected, since the images were binned into four orientation directions for analysis. If the local orientation of the domain is randomly determined and independent from its neighboring domain, we would expect the probability for the four orientation bins to be $\sim 0.25$. This is a good check on our moments analysis, and confirms that the local orientational order is random and globally isotropic.

The average domain size (second moment) $\overline{\langle{s}\rangle}$, scales with the mean correlation length $\overline{\xi}$ with exponent $1.5\pm 0.1$ (Fig. \ref{fig:fig11}(iii), Appendix \ref{appendixB}, TABLE \ref{tab:table17}). The simulation data again fits into this relationship with the experimental data. Using the color-coding for the contour length again shows the underlying dependence, as we observed for the zeroth moment. The magnitudes of the values for the power law of the average number of clusters and the average cluster size are the same, within the uncertainty of the fit. This is reasonable, since we place each pixel into a cluster, so the clusters are space-filling in this analysis. This also serves as a check on our analysis methods.  

\begin{figure*}[!ht]
\includegraphics[scale=0.45]{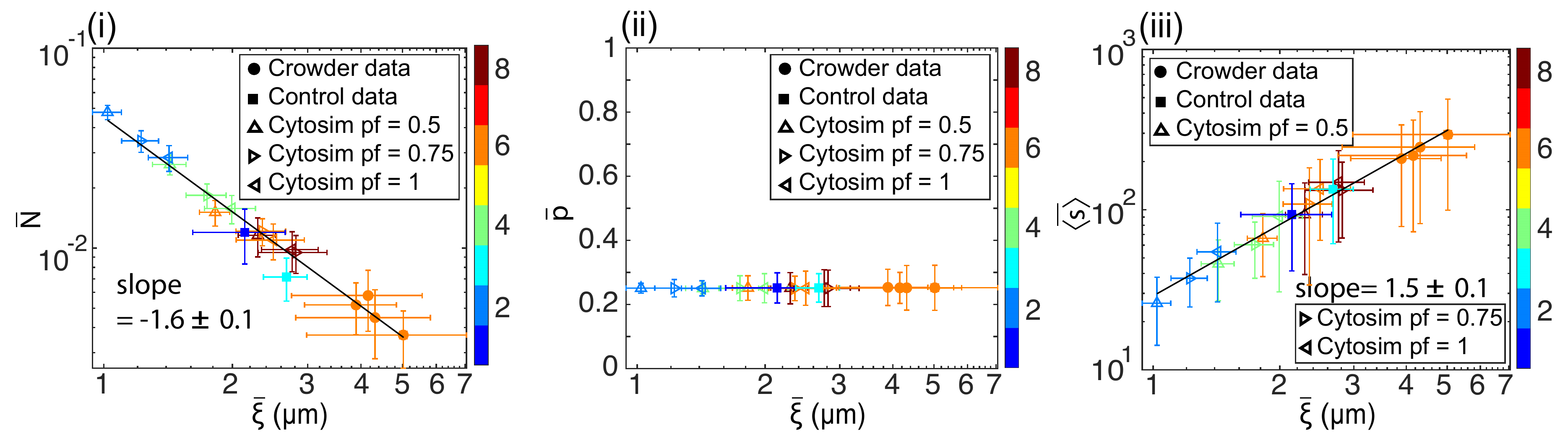}
\caption{\label{fig:fig11}
(i) Average number of m-type (m=1,2,3,4) domains per image $\overline{N}$ plotted over mean correlation length $\overline{\xi}$ in log-log scale. 
(ii) Mean probability of m-type cluster present in an image $\overline{p}$ plotted as a function of $\overline{\xi}$ in semi-log scale. 
(iii) Average cluster size $\overline{\langle s\rangle }$ as a function of $\overline{\xi}$ in log-log scale. Color bar on the right corresponds to the median microtubule length expected for these data sets, as reported before in \cite{Edozie2019Self-organizationStructures}}  
\end{figure*}

Overall, the results indicate that filament length is the most likely underlying control parameter for the surface patterns that we observe. For the data with different crowding agents, we observe only modest differences between the crowders. There are three possible roles played by the crowders in our experiments. First, crowders induce bundling through the depletion interaction and exert compaction force to other filaments. Second, patterns are likely determined by the interaction with the surface, which can be affected by the depletion interaction with filaments and the surface. Indeed, the crowding agent is necessary for any patterns to form on the surface. Third, crowders could influence nucleation and tubulin assembly in initial stages, which would control the average microtubule number $N$  and contour length $l_{c}$. Given the simulation results, we anticipate that this last mechanism is likely the cause for the subtle differences we observe for the average domain size (Fig. \ref{fig:fig2},\ref{fig:fig11}, TABLE \ref{tab:table4}).\\

{\bf Turbidity measurements for different crowders:} 
Our data begs the question: are the subtle differences we observe for different crowders due to the polymers altering the nucleation and growth kinetics to result in slightly different final contour lengths? We cannot directly determine the contour lengths for microtubules in the presence of these crowders using the microscope because the polymers also cause bundling. Prior experiments on tubulin nucleation and growth used a turbidity method and showed that the formation of microtubules was proportional to the PEG concentration for a fixed molecular weight of PEG molecules \cite{herzog1978}. Specifically, higher concentrations of PEG increased the microtubule association constant, $K_{a}$, a determinant of microtubule growth, and reduced the lag time for nucleation. A second study showed that the $K_{a}$ also increased as the PEG size increased \cite{Lee1979InteractionGlycols}.  

\begin{figure}[!ht]
\includegraphics[scale=0.35]{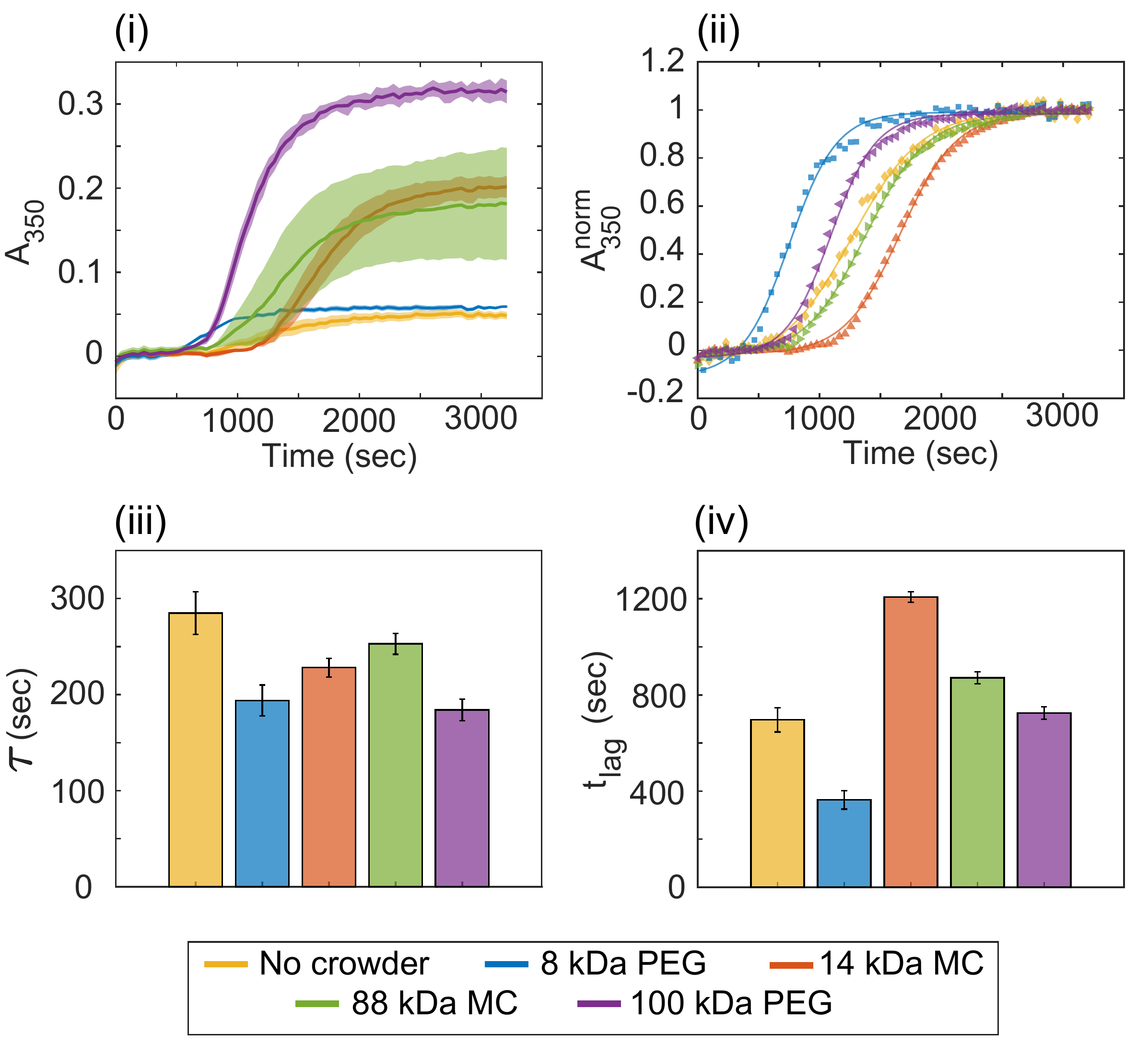}
\caption{\label{fig:fig10} Kinetics of polymerization for 13.6 $\mu$M tubulin: (i)  Absorbance $A_{350}$ data plotted vs. time for no crowder (yellow), 8 kDa PEG (blue), 14 kDa MC (orange), 88 kDa MC (green), and 100 kDa PEG (magenta) with same polymer concentrations used in pattern formation experiments. The shaded regions are the standard deviation resulting from three independent measurements for each configuration (n=3) using MATLAB \cite{mushall2020}. (ii) Normalized absorbance data plotted vs. time no crowder (yellow diamonds), 8 kDa PEG (blue squares), 14 kDa MC (orange upright triangles), 88 kDa MC (green, right pointing triangles), and 100 kDa PEG (magenta left pointing triangles) and fit with Boltzmann sigmoid equation (eq. 1, displayed as same color lines) (Appendix \ref{appendixB}, TABLE \ref{tab:table18}). (iii) Characteristic time $\tau$ is shown in bar plot for the given configurations. Error bars represent 95$\%$ confidence interval for the fit parameter. (iv) Lag time $t_{lag}$, calculated from $\tau$, shown as bar plot for each configuration. Error bars calculated from 95$\%$ confidence interval from the fit.} 
\end{figure}

We notice that the correlation lengths and the average domain sizes appear to be higher for MC molecules than for PEG molecules (Fig. \ref{fig:fig2}, TABLE \ref{tab:table4}). This would imply that PEG molecules could be better at enhancing nucleation and growth of microtubules than MC polymers at the concentrations we use in our study. To directly quantify how the specific crowders we used in this study can affect tubulin nucleation and growth, we perform kinetic growth studies over time using turbidity. First, we compare the raw data without normalization and find that the maximum scattering depends on the crowding agent used (Fig. \ref{fig:fig10}(i)). Typically, the maximum turbidity signal would correlate to the total polymer mass, but there is likely enhanced scattering due to bundling in addition to the increased polymer mass.  

In order to quantify the nucleation and growth time scales in these data, we normalize the data to start at 0 and become maximum at one. We fit the normalized data with a Boltzmann sigmoid equation (\ref{eq:boltzman}) (Fig. \ref{fig:fig10}(ii)). From the fit parameters (Appendix \ref{appendixB}, TABLE \ref{tab:table18}) we find the characteristic timescale for growth, $\tau$, varies with crowders (Fig. \ref{fig:fig10}(iii)). If $\tau$ is small, starting from nascent oligomers it takes less time to polymerize filaments, which will generate more microtubules of shorter length on average. Interestingly, we see the same pattern for the characteristic growth time that we observe and for the correlation length in the inset of Fig. \ref{fig:fig2}C(i). The similarities between these different data imply that the crowders could be changing the nucleation and growth kinetics resulting in the modest pattern changes we observe.

From the turbidity data, we can also quantify the lag time for nucleation (Fig. \ref{fig:fig10}(iv)). 
It appears that the PEG molecules, specifically the 8 kDa PEG, help enhance nucleation while the MC polymers inhibit nucleation. That is an interesting and unexpected result, since macromolecular crowding is expected to help bring objects together in general. One biochemical difference between PEG and MC is that MC is charged. It is perhaps this charge that could be causing interactions with the tubulin dimers to interfere with binding. Future, systematic experiments exploring different polymers with various charges, molecular weights, and concentrations may reveal more information about how polymer characteristics affect microtubule nucleation and growth. \\

\begin{figure}[!ht]
\includegraphics[scale=0.3]{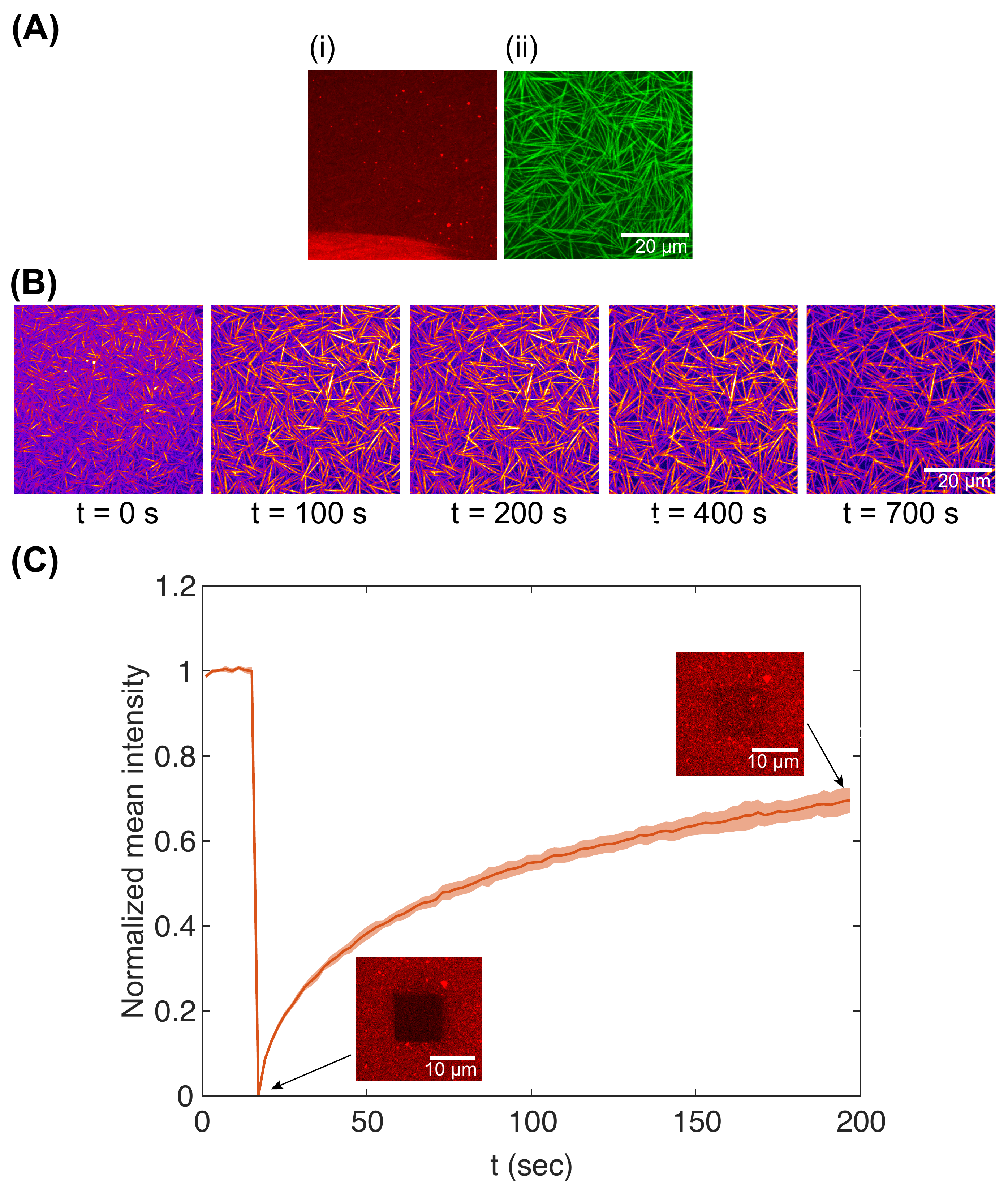}
\caption{\label{fig:fig5} (A) Two color image show the lipid bilayer (561 nm, red) and the disorganized microtubule pattern 488 nm, green). Scale bar is 20 $\mu m$ for both images. (B) Timeseries of the organization on lipid surface in presence of 88 kDa MC. Scale bar is 20 $\mu m$ for all images. (C) Fluorescent recovery after photobleaching of fluorescent lipid bilayer surface demonstrates it is a fluid. Fluorescence intensity recovers 70$\%$ after 180 seconds. Shaded region indicates standard deviation of multiple measurements (n=6). Representative images inset to demonstrate recovery. Scale bar is 10 $\mu m$. } 
\end{figure}

{\bf Lipid surfaces with different crowders:} Because crowding agents cause the depletion of microtubules onto the surface, we wanted to test the effects of the surface coating on microtubule self-organization in the absence of crosslinkers. We replace the polymer brush surface with a lipid bilayer surface to test the surface effects. We are able to recapitulate qualitative fan-like structures on the lipid surface (Fig. \ref{fig:fig4-1}B), although the patterns are not as reproducible as those observed on polymer surfaces. More often, we observe bundles that overlap and are less locally aligned in the presence of the lipid bilayer (Fig. \ref{fig:fig5}A). Indeed, these patterns appear to be isotropic arrangements of bundles, not fan-like organizations. 

We directly observe microtubules nucleating and growing on the surface to determine why the patterns are different (Fig. \ref{fig:fig5}B). When we examine the dynamics of the microtubules forming on the lipid bilayer surfaces, we often observe in initial stage microtubules do not show translation and rotational motion. They grow with time keeping their orientation fixed to overlap or crossover each other to make the pattern less aligned. This is surprising, since the lipid bilayer should behave like a 2D fluid. We monitored the fluidity of the lipid bilayer using fluorescence recovery after photobleaching on a fluorescent lipid bilayer (Fig. \ref{fig:fig5}C). We see recovery of the lipid fluorescence, as expected for a fluid surface. It is possible that the lack of mobility can be due to the interaction with the lipid surface, disrupted depletion giving access to z-dimension. The patterns do not depend on the crowding agent used. So, the interaction between microtubules and the surface play a role in determining whether fan-like structures could be formed and appeared to be more influential than the specific crowding agent. 

To attempt understand the mechanisms that control the patterns on lipid bilayer surfaces, we perform simulations using Cytosim. Our experimental system is quasi-2D due to crowders, so we limit our Cytosim simulation space to 2D plus steric interactions (2D+S) \cite{Rickman2019EffectsSelf-organization}. As shown in TABLE~\ref{tab:table1} the viscosity (dynamic) of the polymer solutions is $\sim$ 0-3 cP (mPa-s) in our experiments. In the simulations we varied the viscosity from 0.0025 to 2.5 Pa-s to determine if the surface effects act as a viscosity parameter (Fig. \ref{fig:fig9}(ii)). 

Although, Cytosim is able to achieve similar fanlike patterns at higher viscosity values (Fig. \ref{fig:fig9}(i)), the dynamics of the pattern formation to reach steady state are distinct from what we observe in experiments. The Cytosim viscosity value for achieving fan-like patterns are 100-1000 fold higher (TABLE \ref{tab:table2}, TABLE \ref{tab:table1}) than the bulk viscosity measurement from the experiments, because we expect that the surface has an additional viscous drag. Indeed, a recent paper from our lab measured the diffusion of single particles near the polymer brush surface and found a 200x reduction in the diffusion coefficient \cite{xu2019}.

%The differences we see in the experiments compared to Cytosim are - on polymer brush surface in initial timescale small microtubules show translation and rotational motion vigorously before they bundle and jam. In the jammed state we observe bundles rotate to align and accommodate themselves inside the pattern. On lipid bilayer surface less oriented `stacked' pattern emerge when we do not observe any translation and rotation. Despite of these differences, Cytosim is able to recapitulate jammed fan-like patterns in the final steady state of the organization.

Overall, the simulations are able to recapitulate the fan-like patterns we observe experimentally on polymer brush surfaces, but the dynamics to achieve the pattern are not the same. For experiments, nucleating and growing bundles were highly mobile on the surface. As they grew, they rotated and settled into their final state. For simulations, high viscosities were needed to achieve fan-like patterns. The nucleating and growing microtubules and bundles were not mobile, and they only rotated upon meeting neighboring filaments. As discussed above, the experiments on lipid bilayer surfaces are not recapitulated well by simulations in the final state or in the dynamics. The lipid bilayer surface is able to disrupt the fanlike arrangement of microtubules, implying that the surface interactions have a significant impact on these beautiful patterns.\\

\subsection{Microtubule organization with MAP65}
In our prior work, we were able to reproducibly form microtubule tactoids in the presence of MAP65 at a concentration of 10$\%$ bound \cite{Edozie2019Self-organizationStructures}. Liquid crystal theory tells us that the competition between elastic deformation energy of the director/microtubule orientation field inside and aniosotropic surface energy at the interface defines the shape of the tactoid and director field inside it. As we showed previously, the tactoids are homogeneous droplets with a constant director field along the major axis of the tactoid \cite{Edozie2019Self-organizationStructures}. Boojums (virtual defects) are located at infinity, not at the poles, as for the bipolar case. We showed that microtubule tactoids are jammed/frozen inside the droplet, demonstrated by imperceptible microtubule turnover when probed using fluorescence recovery after photobleaching (FRAP)\cite{Edozie2019Self-organizationStructures}. FRAP also showed that MAP65 turnover was fast due to the relatively weak binding constant, 1.2 $\mu$M \cite{Edozie2019Self-organizationStructures}. \\

\begin{figure}[!ht]
\includegraphics[scale=0.35]{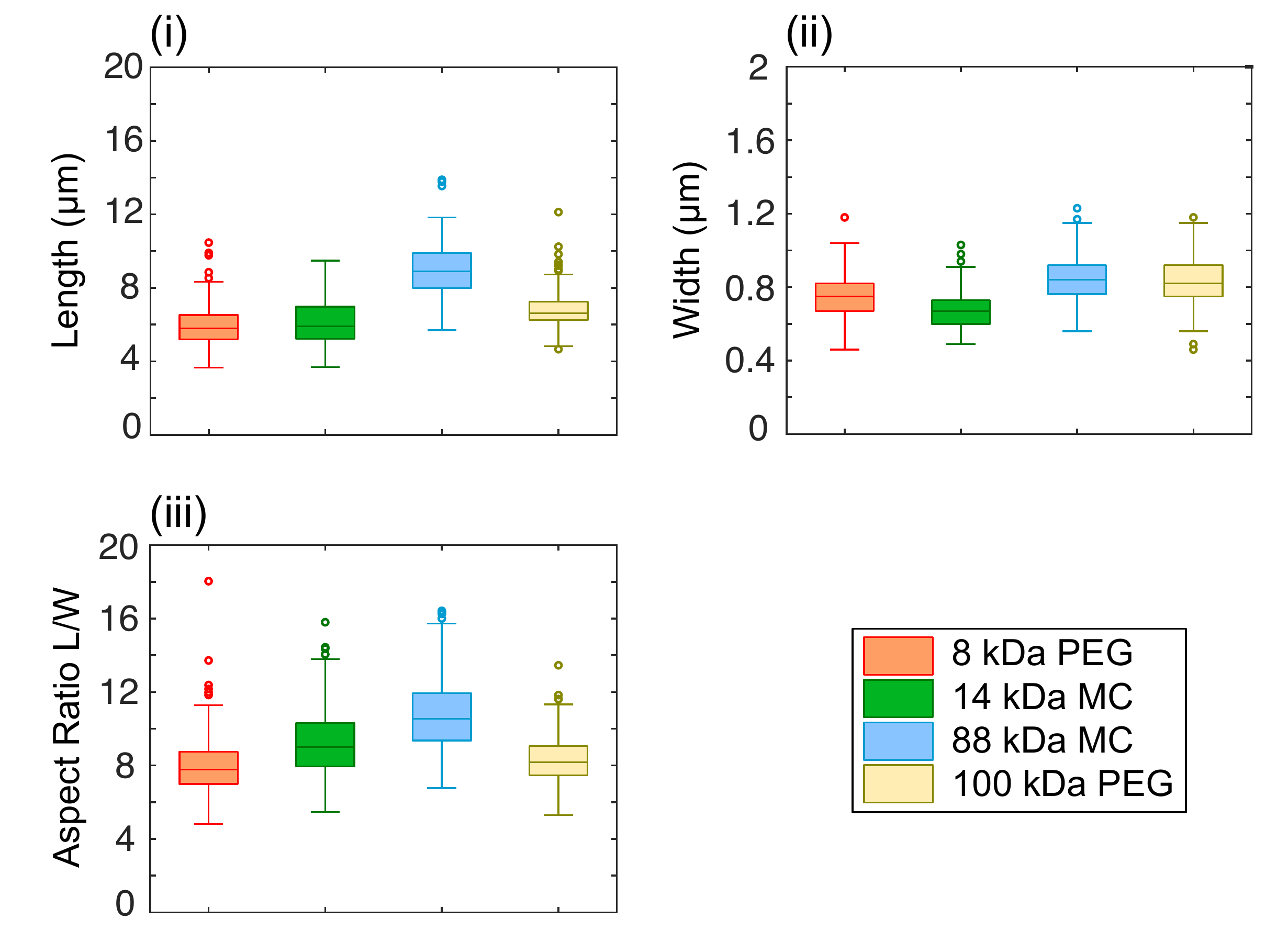}
\caption{\label{fig:fig3} Box-whisker plots of tactoid characteristics with 10$\%$ MAP65 on the polymer brush surface: (i) length of the tactoids, (ii) width of the tactoids, and (iii) aspect ratio of the tactoids. Tactoids are formed in presence of  8 kDa PEG (orange, n = 205), 14 kDa MC (green, n = 217), 88 kDa MC (blue, n = 207), 100 kDa PEG (yellow, n = 240).} 
\end{figure}

{\bf Polymer surface with different crowders:}
Here, we are interested in examining the effect of different crowding agents on the formation, shape, and dynamics of these tactoids. For all experiments, 13.6 $\mu$M tubulin was used with 10\% MAP65 to induce tactoid formation on polymer brush surfaces. In order to quantify the tactoid shape, we use a custom MATLAB code and manual measurements in FIJI, to measure the maximum length, $L=2l$, where $l$ is the half-max length, and width, $W=2r$, where $r$ is the half-max radius of imaged tactoids. We calculate the aspect ratio $L/W$ (TABLE \ref{tab:table3}). Tactoids that form on polymer surfaces in the presence of PEG or MC have subtle differences in their lengths (Fig. \ref{fig:fig3}(i)). The 88 kD MC made slightly longer tactoids than other crowders. Using the KS statistical tests, we find that the length measurements are all statistically different from one another (Appendix \ref{appendixB}, TABLE \ref{tab:table11}). As we observed previously, the widths of the tactoids are in a small range from 500 $\sim$ 1500 nm (Fig. \ref{fig:fig3}(ii)). The 88 kD MC and 100 kD PEG are statistically the same, 8 kD PEG and 14 kD MC are statistically distinct from all other distributions (Appendix \ref{appendixB}, TABLE \ref{tab:table13}). Since the diffraction limit of our microscope is about half the wavelength of the light we image, the widths are only about 2-3 times the diffraction limit, and are less reliable than the length measurements. 

Using the lengths and widths, we calculate the aspect ratio for each crowder. We find that all the distributions are statistically distinct using the KS statistical test, and the data show a dependence on crowder type (Fig. \ref{fig:fig3}(iii), (Appendix \ref{appendixB}, TABLE \ref{tab:table15})). We observe that the aspect ratio of tactoids made in the presence of PEG are always smaller than those made with MC. Interestingly, the dependence of tactoid aspect ratios mirrored the trends we see for the fan-like patterns of self-organization in the absence of MAP65 crosslinker (TABLE \ref{tab:table4}, TABLE \ref{tab:table3}, Fig. \ref{fig:fig2}C(i) inset). Given that we have directly measured that PEG changes the nucleation and growth of microtubules, this seems the most likely mechanism for the altered aspect ratios of these tactoids. In our prior work, we did not examine if the contour length of the microtubules altered the aspect ratio of the tactoids because we could not achieve the same experimental parameters for the highest concentrations of tubulin which gave shorter contour lengths. This new data implies that the nucleation and growth kinetics could impact the final aspect ratio of the tactoids. 
%We previously reported the effects of the crosslinker percentage on the aspect ratio, and found that the aspect ratio decreased with increasing crosslinker \cite{Edozie2019Self-organizationStructures}. Further, MAP65 is known to additionally alter nucleation and growth, which will alter the filament lengths and have an additive effect on the aspect ratio changes \cite{Li2007AtMAP65-1Assembly,Mao2005TwoMicrotubules,Stoppin-Mellet2013MAP65Formation}.
\\

\begin{figure}[!ht]
\includegraphics[scale=0.35]{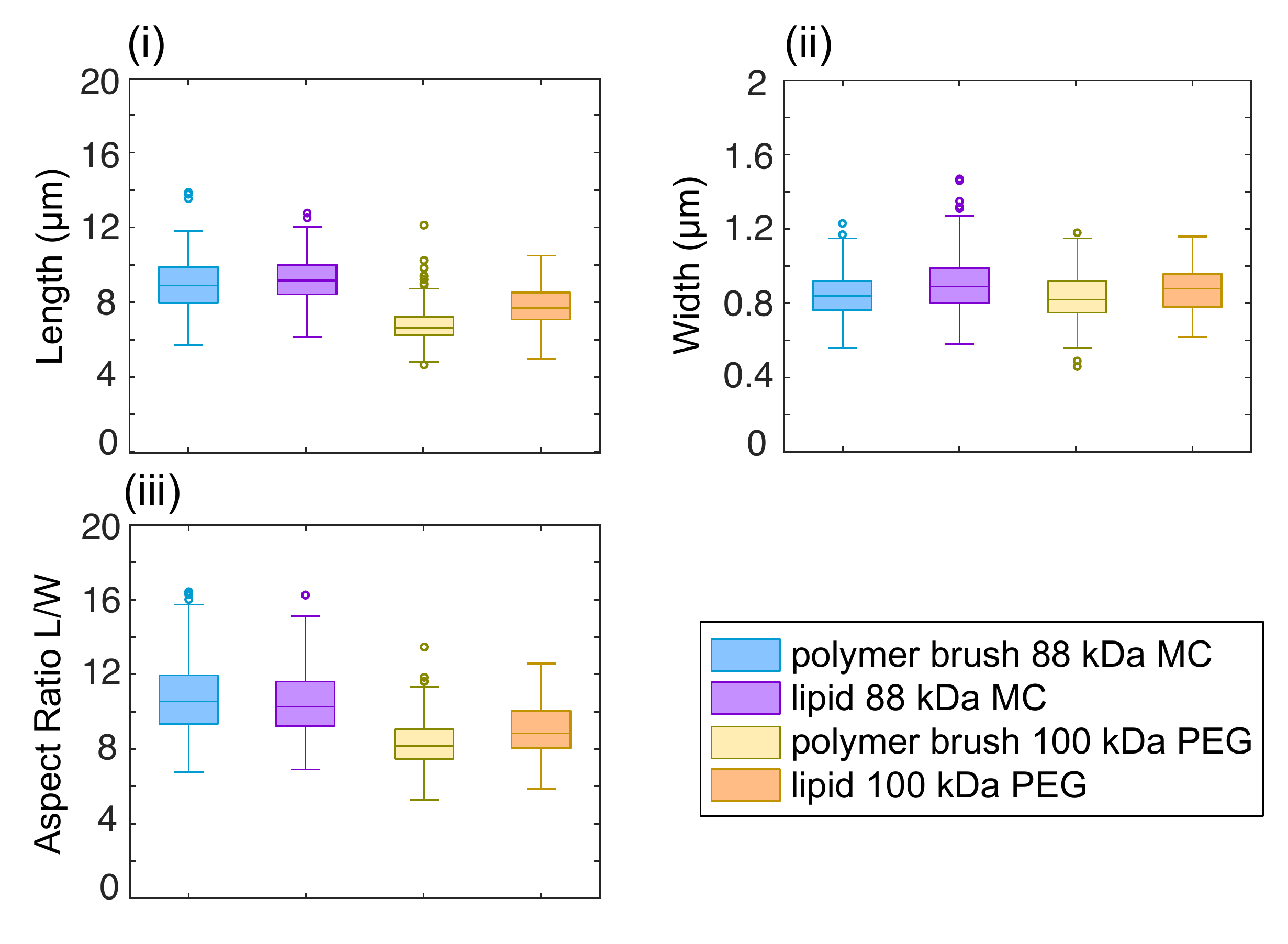}
\caption{\label{fig:fig4-2} Box-and-whisker plots of (i) length, (ii) width, and (iii) aspect ratio of tactoids in presence of two different crowders, 88 kDa MC and 100 kDa PEG, on polymer brush and lipid coated surfaces. For all plots, 88 kDa MC tactoids on polymer brush surface  (light blue, $n=207$), 88 kDa MC tactoids lipid surface  (purple, $n=374$), 100 kDa PEG tactoids on polymer brush surface (yellow, $n=240$) and 100 kDa PEG tactoids on lipid surface (orange, $n=152$).}
\end{figure}

{\bf Lipid surface with different crowders:}
On lipid surfaces, we are able to reproduce tactoids consistently in the presence of different crowders.  %We find that the crowder effects on tactoid length, width, and aspect ratio observed on polymer surfaces are reproducible on lipid surfaces (data not shown). 
We focus on the the effects of the surface conditions for only 88 kDa MC and 100 kDa PEG. For these large polymers, we observe distinct differences in the lengths and aspect ratios comparing the 88 kDa MC and 100 kDa PEG, but not the widths (Fig. \ref{fig:fig4-2}, Appendix \ref{appendixB}, TABLE \ref{tab:table12}, \ref{tab:table14}, \ref{tab:table16}).\\

\begin{table*}[!ht]
\caption{\label{tab:table3}Tactoid features: mean length, mean width, mean aspect ratio and dimensionless anchoring strength. The difference between the top four rows and the bottom four rows is that the data from the top rows were measured by hand using FIJI and the bottom rows were measured using MATLAB. As indicted, the first four columns are for polymer brush surfaces and last two columns are for lipid surfaces.}
\begin{ruledtabular}
\begin{tabular}{ccccccc}
 &\multicolumn{4}{c}{Polymer brush surface}&\multicolumn{2}{c}{Lipid surface}\\
 \textrm{Crowders}&\textrm{8 kDa}&\textrm{14 kDa}&\textrm{88 kDa}&\textrm{100 kDa}&\textrm{88 kDa}&\textrm{100 kDa}\\
\textrm{}&\textrm{PEG}&\textrm{MC}&\textrm{MC}&\textrm{PEG}&\textrm{MC}&\textrm{PEG}\\ 
\textrm{}&\textrm{(mean $\pm$ SD)}&\textrm{(mean $\pm$ SD)}&\textrm{(mean$\pm$ SD)}&\textrm{(mean$\pm$ SD)}&\textrm{(mean $\pm$ SD)}&\textrm{(mean $\pm$ SD)}\\ 

\hline
L($\mu m$) & 5.9$\pm$ 1.0 & 6.1$\pm$ 1.2 & 8.9$\pm$ 1.4 & 6.8$\pm$ 0.9 & 9.2$\pm$ 1.1 & 7.7$\pm$ 1.0\\
W($\mu m$) & 0.8$\pm$ 0.1 & 0.7$\pm$ 0.1 & 0.8$\pm$0.1 & 0.8$\pm$ 0.1 & 0.9$\pm$ 0.1 & 0.9$\pm$ 0.1\\
L/W & 8$\pm$1.7 & 9.2$\pm$1.7 & 10.7$\pm$1.9 & 8.3$\pm$1.3 & 10.4$\pm$1.7 & 8.9$\pm$1.4\\
$\omega \sim$ & 16 & 21 & 28 & 17 & 27 & 20 \\ 
\hline
L($\mu m$) & 5.6$\pm$ 0.9 & 5.6$\pm$ 1.1 & 8.8$\pm$ 1.5 & 6.2$\pm$ 0.9 & 8.8$\pm$ 1.2 & 7.2$\pm$ 1.0\\
W($\mu m$) & 0.9$\pm$ 0.1 & 0.8$\pm$ 0.1 & 1.0$\pm$0.1 & 0.9$\pm$ 0.1 & 1.0$\pm$ 0.1 & 0.9$\pm$ 0.1\\
L/W & 6.3$\pm$1.1 & 7.0$\pm$1.5 & 9.0$\pm$1.9 & 6.7$\pm$1.0 & 9.3$\pm$1.3 & 7.8$\pm$1.4\\
$\omega \sim$ & 10 & 12 & 20 & 11 & 22 & 15\\
\end{tabular}
\end{ruledtabular}
\end{table*}

\begin{figure*}[!ht]
\includegraphics[scale=0.5]{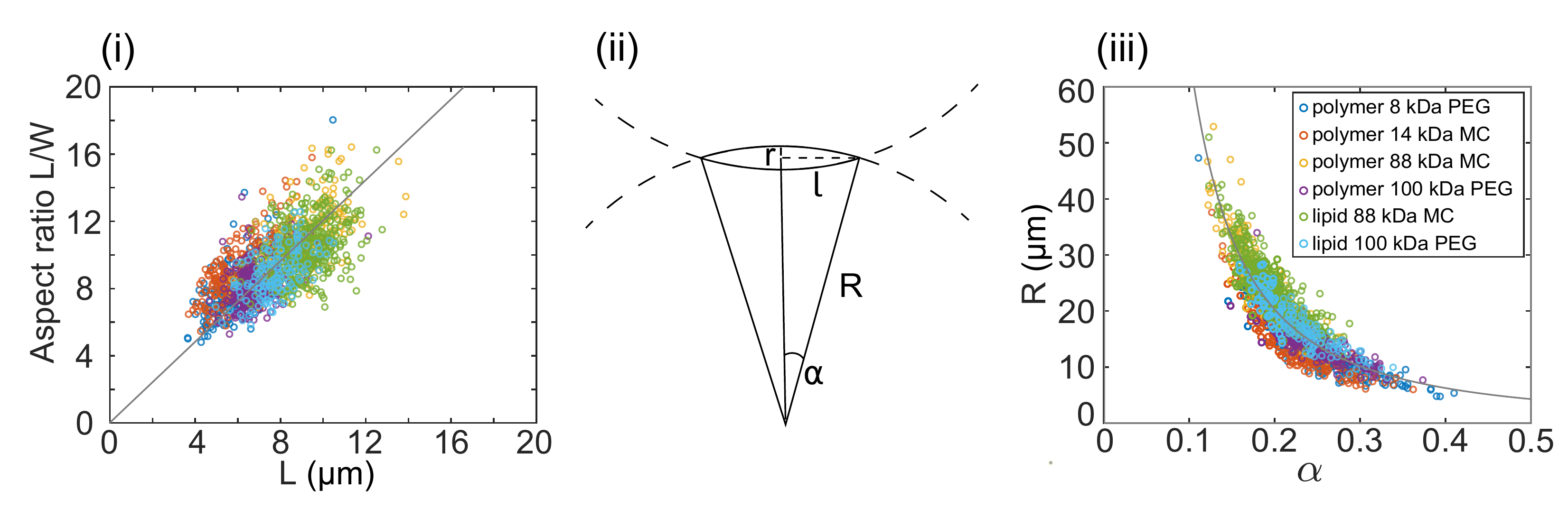}
\caption{\label{fig:fig8} Tactoid shape characteristics for all presented data. (i) Plot of tactoid aspect ratio vs tactoid length $L=2l$ shows a slope equivalent to the constant width of the tactoids. (ii) Diagram shows the relationship between quantitative parameters of the tactoid $r$, $l$, $R$ and $\alpha$. (iii) Plot of $R$ vs $\alpha$ demonstrates a power law behavior due to the constant tactoid width. The gray lines in (i) and (iii) represent fits to the data (Appendix \ref{appendixB}, TABLE \ref{tab:table20}, \ref{tab:table21})).}
\end{figure*}

{\bf Microtubule tactoid shapes cannot be explained by liquid crystal theory:}
To compare our tactoid results to liquid crystal theory and attempt to elucidate the physical mechanisms controlling tactoid formation, we use the length, width, and aspect ratio measurements from all the data to examine the shape parameters (Fig. \ref{fig:fig8}). We plot the aspect ratio $L/W$ as a function of $L$, and find that the dependence appeared linear with a positive slope, $m=1.21\pm0.01$ (Fig. \ref{fig:fig8}(i), Appendix \ref{appendixB}, TABLE \ref{tab:table20}). The linear dependence indicates the constant width of the tactoids. The plot has a large scatter, which is due to the wide range of data for the lengths and widths (Fig. \ref{fig:fig8}(i)). Despite the scatter, this result suggests that the width is constant for all tactoids, regardless of the crowding agent and surface used (Fig. \ref{fig:fig8}(i)). Prior work on other systems of homogeneous tactoids have not observed any dependence of the aspect ratio on the length of the tactoids. For instance, homogeneous tactoids made from carbon nanotubes have been shown experimentally to display a constant aspect ratio as a function of tactoid length, implying that the length and width grew together \cite{Puech2010NematicNanotubes}. The difference between prior systems and the microtubule system could be that our system is driven to condense using specific crosslinkers and is not fluid-like. 

We can examine tactoid shape in another way, using two additional variables to characterize the tactoid: $R$ and $\alpha$. These parameters come from thinking of the tactoid surface as an arc of a circle with radius $R$ making an angle $2\alpha$ at the center (Fig. \ref{fig:fig8}(ii)). These shape parameters can be related to the semi-major length $l=L/2$ and and semi-minor length $r=W/2$, as so: 

\begin{equation}
R=\frac{l^{2}+r^{2}}{2r}
\end{equation}
\begin{equation}
\alpha=\sin^{-1} \left( \frac{2lr}{l^{2}+r^{2}}\right)
\end{equation}

We find that when we plot $R$ as a function of $\alpha$ we see an inverse dependence. This dependence is another indicator that the width is constant in our system. When we fit the data for $R$ vs $\alpha$ with a power law, it yields an exponent $\delta=-1.69\pm 0.04$ (Appendix \ref{appendixB}, TABLE \ref{tab:table21}). This behavior is the opposite of what was seen previously in case of actin tactoid condensates \cite{Oakes2007GrowthF-actin}. 

In the prior work, the actin tactoids were bipolar, so that the $R$ dependence on $\alpha$ could be used to determine the rescaled Frank free energy constants for splay $\kappa_{1}$ and bending $\kappa_{3}$ \cite{Oakes2007GrowthF-actin}.  Bipolar tactoids are described by the elastic energy terms as in eq. (\ref{A1}) (see Appendix \ref{appendixC} for details). Although our tactoids are not bipolar, if we were to use the same assumptions to fit our results, we would end up with a negative bending constant, which is nonphysical. 

For homogeneous tactoids, which our microtubule tactoids appear to be, the surface energy term in free energy should dominate the energy landscape (eq. (\ref{A2})). Further, elastic terms are negligible because there are no distortions in the director field. In liquid crystal theory using scaling arguments and minimization of free energy via variational theory, it has has been shown that $\omega$, the ratio of the anchoring strength and the bare surface tension, is the sole parameter describing systems with homogeneous tactoids \cite{Prinsen2003ShapeTactoids}. We calculated $\omega$ values using $L/W=2\omega^{1/2}$ for $\omega \gg 1$. In our case, $L/W$ is much larger than one, as seen in  TABLE \ref{tab:table3}. Using the average value of $L/W$ for our systems, we find that $\omega$ ranges from 16 - 28. These $\omega$ values are large compared to other systems \cite{Puech2010NematicNanotubes, Bagnani2019AmyloidTactoids, Weirich2017LiquidBundles}. 

Our tactoids are different from other system in a few ways. First, being a homogeneous tactoid, the aspect ratio does not stay constant with the length of the tactoid as would be predicted from liquid crystal theories and experiments. Second, the $\omega$ value is  high although the tactoids are homogeneous. In our prior work, we demonstrated that microtubule tactoids are immobile and likely jammed within the tactoid \cite{Edozie2019Self-organizationStructures}. Since the microtubules within the tactoids do not move, they do not behave like liquid droplets. Microtubule tactoid condensates might be more reminiscent of solid condensates in a liquid background - like the precipitation of crystals. \\

\begin{figure}[!ht]
\includegraphics[scale=0.35]{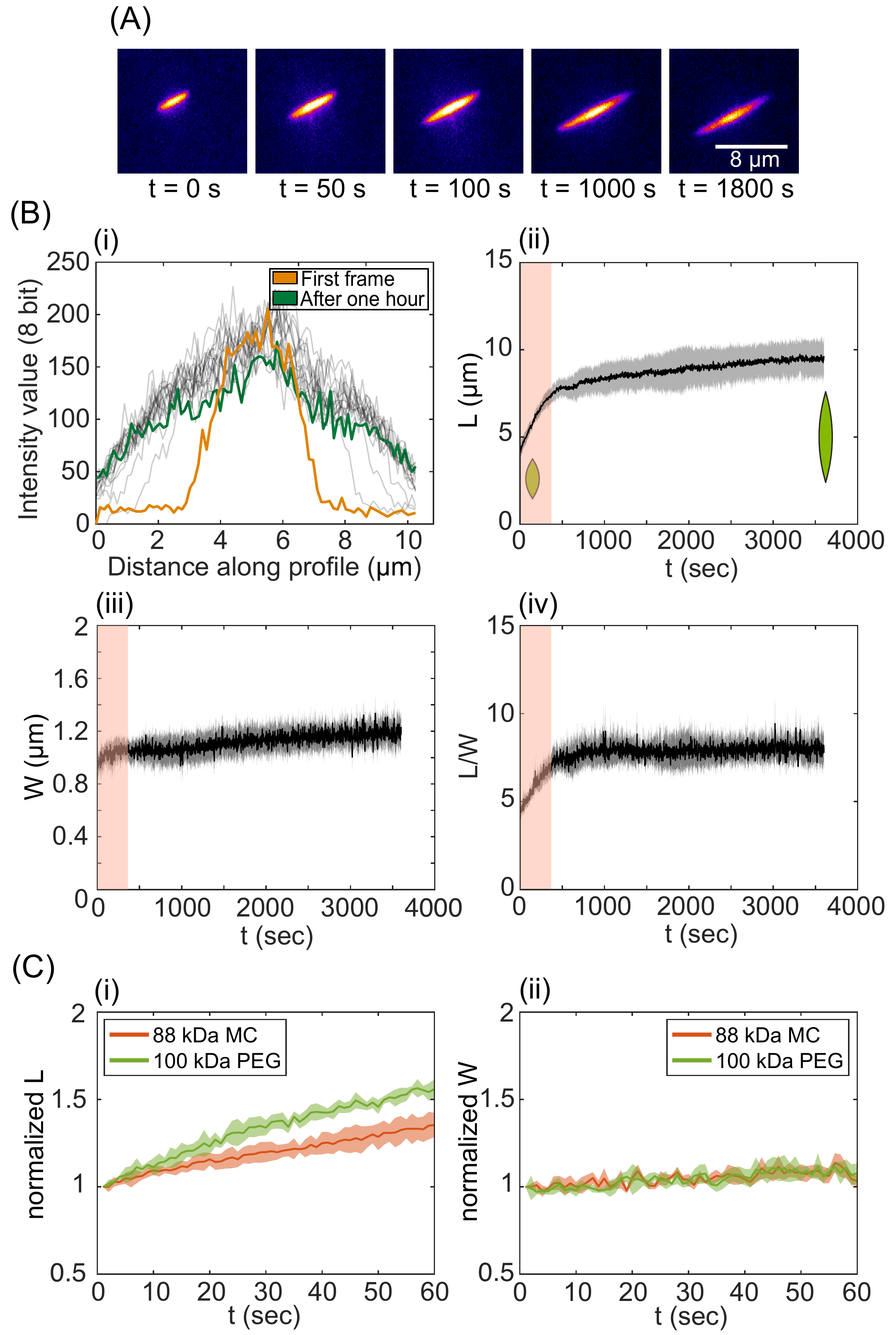}
\caption{\label{fig:dynamics} (A) Timeseries of a single tactoid growth on lipid surface in presence of 100 kDa PEG as crowder. Here t = 0s represents the starting point of data acquisition. (B) Tactoid growth characterization: (i) Intensity distribution along tactoid length for the first frame (orange line), over time (gray lines) and after one hour (green line) is shown. (ii) Tactoid length L is plotted over time. (iii) Tactoid width W is plotted over time, (iv) Tactoid aspect ratio, L/W, is plotted over time. For (ii)-(iv), three tactoids from the same video are averaged. Data is taken in the presence of 88 kDa MC on lipid surface. (C) Tactoid growth rate comparison in presence of two crowders 88 kDa MC (red shaded curve, n=4) and 100 kDa PEG (green shaded curve, n=3) on lipid surface within first minute (i) normalized tactoid length L and (ii) normalized tactoid width W. The shaded regions indicate the standard deviation of the mean.  }
\end{figure}

{\bf Tactoid Growth Dynamics:} In order to help understand the mechanism of tactoid formation, we directly observe how a single tactoid nucleated and grew in time by quantifying the intensity profile as a function of time (Fig. \ref{fig:dynamics}A). We plot only the intensity profile of the major axis over time as an example of the analysis (Fig. \ref{fig:dynamics}B(i)). When we plot the length of three tactoids as a function of time, we find that the tactoid growth displayed two distinct growth phases (Fig. \ref{fig:dynamics}B(ii)). In early times, the tactoid nucleates and grows quickly. The tactoid is short, likely having formed from sub-resolution, highly mobile microtubule filaments associating via the crosslinking protein. The tactoid length increases rapidly at a rate $v_{t} = 10.7\pm0.2$ nm/s (Appendix \ref{appendixB}, TABLE \ref{tab:table19}), fit from $t = 0$s until $t = 200$s. This tactoid length growth rate suggests that this initial growth is because of microtubule elongation and not because of new microtubule addition, as this growth rate is one third of the microtubule growth rate in this tubulin concentration \cite{Rickman3427}.  The time denoted $t = 0$ s is the time when chamber is placed on the microscope which was $\sim$ one minute after the elongation mix is flowed through the chamber. At longer times, the tactoid length either reaches a stable state or shows very little growth with a slope $0.39\pm0.02$ nm/s (Appendix \ref{appendixB}, TABLE \ref{tab:table19}). 

Interestingly, the width of these initial assemblies is already in the range of 500 - 1500 nm, the same range we observe for all tactoids (Fig. \ref{fig:fig3}(ii), \ref{fig:fig4-2}(ii)). This data implies that the width of the tactoids is set very early in nucleation and growth, and changes little over time. We observe a small change in width over time, with a slope of  $0.048\pm0.002$ nm/s (Fig. \ref{fig:dynamics}B(iii), Appendix \ref{appendixB}, TABLE \ref{tab:table19}). 

%The aspect ratio trend follows the length growth trend, as expected. The long-time phase shows a constant aspect ratio over time for all three tactoids. Of the three we measured, two continued a slow growth in both length and width, while the last one is constant in length and width. The constant shape parameters imply that the tactoids have reached equilibrium likely due to exhausting the tubulin dimer supply.

The aspect ratio $L/W$ also has two phases as a function of time (Fig. \ref{fig:dynamics}B(iv)). Since the width does not change much, even in the initial fast-growing phase, the aspect ratio's trend follows that of the length, with an initial fast growing rate of $r_{t} = 0.0088\pm0.0006 ~s^{-1}$ (Appendix \ref{appendixB}, TABLE \ref{tab:table19}). At the later phase, the aspect ratio does not change and has a constant aspect ratio of $L/W = 7.7 \pm 0.1$ (Appendix \ref{appendixB}, TABLE \ref{tab:table19}), which we find by fitting the data from t=2000s to t = 3600s to a line and determining the offset.

We also observe a change in tactoid length growth rate in initial times in presence of different crowders. In presence of 100 kDa PEG we observe faster tactoid length growth than tactoids in presence of 88 kDa PEG. We have normalized the tactoid length to observe this difference as shown in Fig. \ref{fig:dynamics}C(i). Initial tactoid length growth is likely due to microtubule elongation. Once again, our observation bolster the conclusion that the crowders affect the growth rate. The trend is similar to Fig. \ref{fig:fig10}(iii) which suggests crowders affect the growth timescale and that is reflected in length difference. The crowders do not cause variation in the tactoid width (Fig. \ref{fig:dynamics}C(ii)).\\

%paragraph that discusses this line 501 stuff up here)

Overall, when examining the tactoids and comparing to prior liquid crystal theory and experiment, we find that the microtubule tactoids are unique. They have solid-like properties with little rearrangement inside themselves. The microtubule tactoid system did not display the homogeneous to bipolar transition. In other entropically driven systems it has been predicted and shown that this transition happens when the size or the volume of the tactoid crosses a threshold value, which depends on $\kappa$ from elastic terms and $\omega$ from surface terms. The homogeneous to bipolar transition was observed experimentally in the CNT system \cite{Jamali2015ExperimentalTactoids} and amyloid fibril systems \cite{Bagnani2019AmyloidTactoids}.  

The primary physical aspect of microtubule tactoids that make in inaccessible to liquid crystal theory is that the tactoid width is approximately constant in all conditions. Due to the limited width, the aspect ratio depends entirely on tactoid length. Further, the width was unaffected by experimental parameters such as the crowding agent or the surface of the chamber.

There are several possible reasons why the width is constant. First, there is a crosslinker causing condensation of the tactoids instead of being entropically driven, like previous systems \cite{Oakes2007GrowthF-actin, Bagnani2019AmyloidTactoids, Jamali2015ExperimentalTactoids}. There was a prior report of actin filament tactoid formation driven by an actin-associated protein crosslinker that were bipolar, behaved as liquids with fluid interiors, and were capable of coalescence \cite{Weirich2017LiquidBundles,Weirich2019Self-organizingDroplets}. The differences between the microtubule and actin tactoids condensed with crosslinkers could be caused by the crosslinker affinity, flexibility, or filament length control which were different for each of these systems. Future studies using engineered MAP65 crosslinkers with altered physical characteristics or binding affinities could begin to probe these possibilities.

%reason1
Another mechanism could be due to the presumably altered affinity of the crosslinker when bound. The equilibrium dissociation constant for MAP65 to bind to single microtubules is $K_{D}\sim 1.2~ \mu M$, indicating weak affinity and a fast turn-over rate \cite{Tulin2012Single-moleculeBundling}. Despite the relatively weak binding, MAP65 can immobilize the microtubules within the tactoid \cite{Edozie2019Self-organizationStructures} and even slow microtubule gliding driven by strong kinesin-1 motoros \cite{Pringle2013MicrotubuleMAP65}. Other works have also demonstrated that MAP65 and its analogs have a higher affinity for microtubules that are crosslinked in an antiparallel manner compared to its affinity on a single filament which can cause compaction forces to shorten the overlapping bundled regions between microtubules \cite{Lansky2015DiffusibleNetworks, Braun2011AdaptiveApart, Subramanian2010InsightsProtein, Pringle2013MicrotubuleMAP65}. If there is strong binding within the bundle, that could control the director field inside the tactoid and result in $\kappa > \omega$. For future models, adding a specific `crosslinking' term in the free energy expression could be used to describe this system more accurately.

An additional consequence of altered affinity of MAP65 when bound is that the concentration outside of the tactoids would becomes depleted compared to the concentration inside where the MAP65 should be sequestered. If a new microtubule or tactoid tries to bind to the already bundled filaments, there could be insufficient available MAP65 to allow binding or coalescence.  Further, already nucleated and polymerized microtubules are slower to diffuse and would not be able to join tactoids easily compared to free dimers. Such new microtubules are unlikely to be nucleated once significant growth has started, due to the depletion of dimers from the environment, so it is more likely for the length to continue to grow than the width. Thus, the width limit could be a natural consequence of the nucleation and growth kinetics of microtubules.

%reason2
Finally, the fixed width could be explained by other physical phenomena. Limited width in assemblies has been explored theoretically by kinetically arrested aggregation models \cite{Lai2007EvolutionBundles}, self-assembly models with long-range interaction schemes \cite{Dutta2016BundleInteractions}, as well as chirality and geometrical packing frustration problems \cite{greggrason2020}. Microtubules have an intrinsic chirality present in their lattice, most clearly shown at the lattice `seam'. If MAP65 binding templates off the microtubule lattice, the tactoids should be chiral bundles as well. This inherent chirality could result in packing frustration which limits the width in the assembly \cite{greggrason2020, hagan2020equilibrium}. 

In support of the idea of a chiral tactoid, prior work on the structure of mitotic spindles has shown that the overlapping microtubules, which are coated with MAP65-like proteins, display a helical twist of the filaments of the spindle \cite{Novak2018TheBundles}. Cross-sectional studies of microtubule bundles in the presence of crowders have shown that increasing depletion can alter the packing from square to hexagonal packing \cite{Needleman2004SynchrotronInteractions, Needleman2005RadialProteins}, but they did not examine if there was chirality. Similar high resolution studies using cryo-electron microscopy on microtubule tactoids have yet to be performed, but they may show a chirality and square packing to accommodate the anti-parallel preference between filaments.  \\

\section{\label{sec:level4} Conclusions}

%What we did and why:
Using bottom-up reconstitution experiments, we examine how entropic forces can control pattern formation in a microtubule system with the without crosslinkers.  We find that the quantitative metrics of the organization subtly depend on the crowder. We are able to perform new simulations in the absence of crosslinkers that show the quantitative metrics of the pattern are most likely controlled by the contour length. The contour length of microtubules in our system is a direct consequence of the nucleation and growth kinetics. We test if the crowding agents alter the nucleation and growth kinetics, and we find the same trends in growth rates as observed in the pattern metrics. These results imply that the crowding effects are likely only changing due to altered contour lengths from altered growth kinetics. We also test the pattern formation in the presence of the crosslinkers to create tactoids and find the same effects due to different crowding agents, again implying that the nucleation and growth kinetics affect the tactoids metrics.

%to conclude that the contour length is the most likely control parameter for the quantitative changes we observe. That Focusing on four different crowding agents with different types, sizes, and concentrations we demonstrated that self-organization patterns subtly depend on crowding agents and surface coatings with or without crosslinkers. 

%For experiments in the absence of specific crosslinkers, microtubules robustly create fan-like patterns in the presence of any crowder. Crowders had subtle effects on the filament nucleation and growth, resulting in changes in the contour length, and ultimately the size of the surface patterns. The fan-like organizations were reproducible in simulations. We found a scaling relation between domain size and correlation length to describe these surface patterns.  

We test if the surface treatment affected the pattern with and without crosslinkers. We find that the surface pattern without crosslinkers is less reproducible on lipid bilayer surfaces despite the increased fluidity of the surface. Tactoid shape metrics are unaffected by the presence of the lipid bilayer. 

%When trying to reproduce this effect by lowering the viscosity in simulations, we did observe some similarities, but the final patterns and evolution of the system in time did not match the patterns observed on lipid surfaces.

%In the presence of antiparallel microtubule crosslinkers, all systems result in tactoid shapes. There was a subtle difference on the tactoid length, following the same trend what we observed for the fan-like patterns. This is further evidence that the crowders are likely affecting microtubule contour length through changes in nucleation and growth of the filaments. 

Using our quantitative metrics, we test our tactoids against existing liquid crystal theories. The tactoids are not comparable to liquid crystal theory because the condensed microtubule phase is not liquid, but appears solid. In addition, the width of the condensed homogeneous tactoids are constant regardless of crowders. 

In this study, the only dynamic component is microtubule growth, which is halted when the free tubulin was exhausted. The mitotic spindle and other biological condensed tactoid systems are fluid and display steady state dynamics. In order to capture these exciting activities, we propose future experiments should include active components, such as motor proteins. Indeed, such experiments have resulted in active nematics of microtubule bundles, asters, and vortices \cite{Sanchez2012SpontaneousMatter,strubing2020, Sanchez2011Cilia-likeBundles,Roostalu2018DeterminantsMotors,Nedelec1997Self-organizationMotors,lemma2019,strubing2020}. Other possible directions to explore to increase the fluidity of these condensates include adding enzymes, such as depolymerizing kinesins or microtubule severing enzymes, to limit microtubule length. Moreover, addition of nucleating centers to restrict nucleation and growth, or crosslinker affinity alteration, or addition of different ionic species are future directions to explore. \\

\section{\label{sec:level5} Acknowledgments}
Experiments were performed by SS, LH, and RQ. Initial analysis was performed by LH and RQ. SS and JLR composed the original manuscript. SS, LH, RQ and JLR edited the manuscript. This work was funded by a grant from the National Science Foundation NSF BIO-1817926 and partially funded from the Keck Foundation to Dr. R. Robertson-Anderson, M. Das, M. Rust, and JLR.

\appendix

\section{Crowder Characterization}\label{appendixA}

In our experiments, two types of crowders (methylcellulose and poly-ethylene glycol) are added to the system, with different molecular weights, and $\%$(w/v) concentrations to modulate the depletion forces.  In our experiments, we only use dilute polymer solutions where polymers are  hydrodynamically separated and cannot interact with each other. Once the concentration of polymer, $c$, crosses the critical overlap concentration $c^{*}$, polymers become entangled and overlapped. In the dilute regime $c<c^{*}$, these blob configurations are characterized by the radius of gyration $R_{g}$. A scaling relation, $R_{g}=K_{R}M_{W}^{\nu}$, connects $R_{g}$ with weight-average molecular weight $M_{W}$, where $\nu$ is the Flory exponent, and $K_{R}$ is an empirical constant of proportionality. Similarly, the intrinsic viscosity,$[\eta] = \lim_{c\to 0}\frac{\eta -\eta_{0}}{c \eta_{0}}$ is related to $M_{W}$ by  Mark-Howink-Sakurada(MHS) equation, $[\eta]=K_{\eta}M_{W}^{\alpha}$, where c is the solute concentration, $\alpha$ is MHS exponent, $K_{\eta}$ is the empirical constant of proportionality, and $\eta,~\eta_{0}$  are the solution, solvent viscosity respectively. Polynomial expression of osmotic pressure $\Pi$ is obtained via virial expansion in terms of molar concentration $C$ ~\cite{Cohen1997AnData.},
\begin{equation}
\Pi=RT[C+aC^{2}+bC^{3}+...]
\end{equation}
where $R$ is universal gas constant and T is temperature in K, and a and b are virial coefficients. For low polymer concentration like our experiment, the lower bound can be well estimated by setting $\Pi\sim CRT$ ignoring higher order terms.\\

\begin{table}[!ht]
\caption{\label{tab:table1}
Crowder characteristics.
}
\centering
\begin{ruledtabular}
\begin{tabular}{c c c c c}
\multicolumn{1}{c}
\textrm{Crowders}&\textrm{8 kDa}&\textrm{14 kDa}&\textrm{88 kDa}&\textrm{100 kDa}\\
\textrm{}&\textrm{PEG}&\textrm{MC}&\textrm{MC}&\textrm{PEG}\\
\colrule

$K_{R}$ (nm) & 0.021 \footnotemark[3]& 0.054 \footnotemark[1]& 0.054 \footnotemark[1]& 0.021\footnotemark[3]\\
$\nu$ & 0.583 \footnotemark[3]& 0.576\footnotemark[1] & 0.576\footnotemark[1] & 0.583\footnotemark[3]\\
$K_{\eta}$ (ml/g)& 0.0488 \footnotemark[2]& 0.102 \footnotemark[1]& 0.102\footnotemark[1] & 0.0064\footnotemark[2]\\
$\alpha$ & 0.66 \footnotemark[2]& 0.704\footnotemark[1] & 0.704 \footnotemark[1]& 0.82\footnotemark[2]\\
$R_{g}$ (nm) $\sim$ & 4 \footnotemark[3]& 13 \footnotemark[1]& 38 \footnotemark[1]& 17\footnotemark[3]\\
$\eta_{ref}$ (cP) & 0.91 & 0.85 & 1.13 & 0.93\\
$\eta_{exp}$ (cP) & 0.97 & 0.96 & 2.13 & 0.98\\
%\colrule
$c^{*}\%$ (w/v) & 4.7 & 1.21 & 0.33 & 0.72\\
c$\%$ (w/v) & 1 & 0.12 & 0.15 & 0.25\\
c in $\mu M$ & 1250 & 85 & 17 & 25\\
$\Pi~(\frac{N}{m^{2}})~>$ & 3221 & 219 & 43 & 64\\

\end{tabular}
\end{ruledtabular}
\footnotetext[1]{ref. \cite{Funami2007ThermalWeights}}
\footnotetext[2]{ref. \cite{brandrup1989polymer}}
\footnotetext[3]{ref. \cite{Devanand1991AsymptoticOxide, ziebacz2011}}
\end{table}

For each polymer, the physical properties such as viscosity $\eta$, radius of gyration $R_{g}$, the critical entanglement concentration $c^{*}$ are  estimated and shown in TABLE \ref{tab:table1}. Polyethylene glycols (PEGs) are a highly soluble, linear, inert polymers, whereas methylcellulose (MCs) are negatively charged, branched, polymers that strongly influence the macroscopic viscosity. For MCs, our parameters are estimated from ref. \cite{Funami2007ThermalWeights}. The $R_{g}$ and $M_{W}$ are insensitive to temperature within the range 15 $^{o}$C - 60 $^{o}$C whereas $\eta$ depends on temperature, and were performed at 15 $^{o}$C and MC concentrations of 0.03$\%$ (w/v) (dilute regime) in the referenced article \cite{Funami2007ThermalWeights}.  For aqueous PEG solution at 35 $^{o}$C, the viscosity parameters are estimated from data in \cite{brandrup1989polymer}. The $R_{g}$ value and $c^{*}=\frac{M_{W}}{\frac{4}{3}\pi R_{g}^{3}N_{A}}$, where $N_{A}$ is Avogadro's number, are extrapolated for our $M_{W}$ from \cite{Devanand1991AsymptoticOxide, ziebacz2011}. We estimate the viscosity, $\eta_{ref}$ from the given values of $K_{\eta}$ and $\alpha$ in the references \cite{Funami2007ThermalWeights}-\cite{ziebacz2011}. In addition, since our experiments are performed at 37 $^{o}$C to enhance microtubule growth, we also measure the viscosity independently $\eta_{exp}$ in TABLE \ref{tab:table1}. \\

\section{Cytosim depletion vs no depletion comparison}\label{appendixAA}
  
\begin{figure}[!ht]
\includegraphics[scale=0.33]{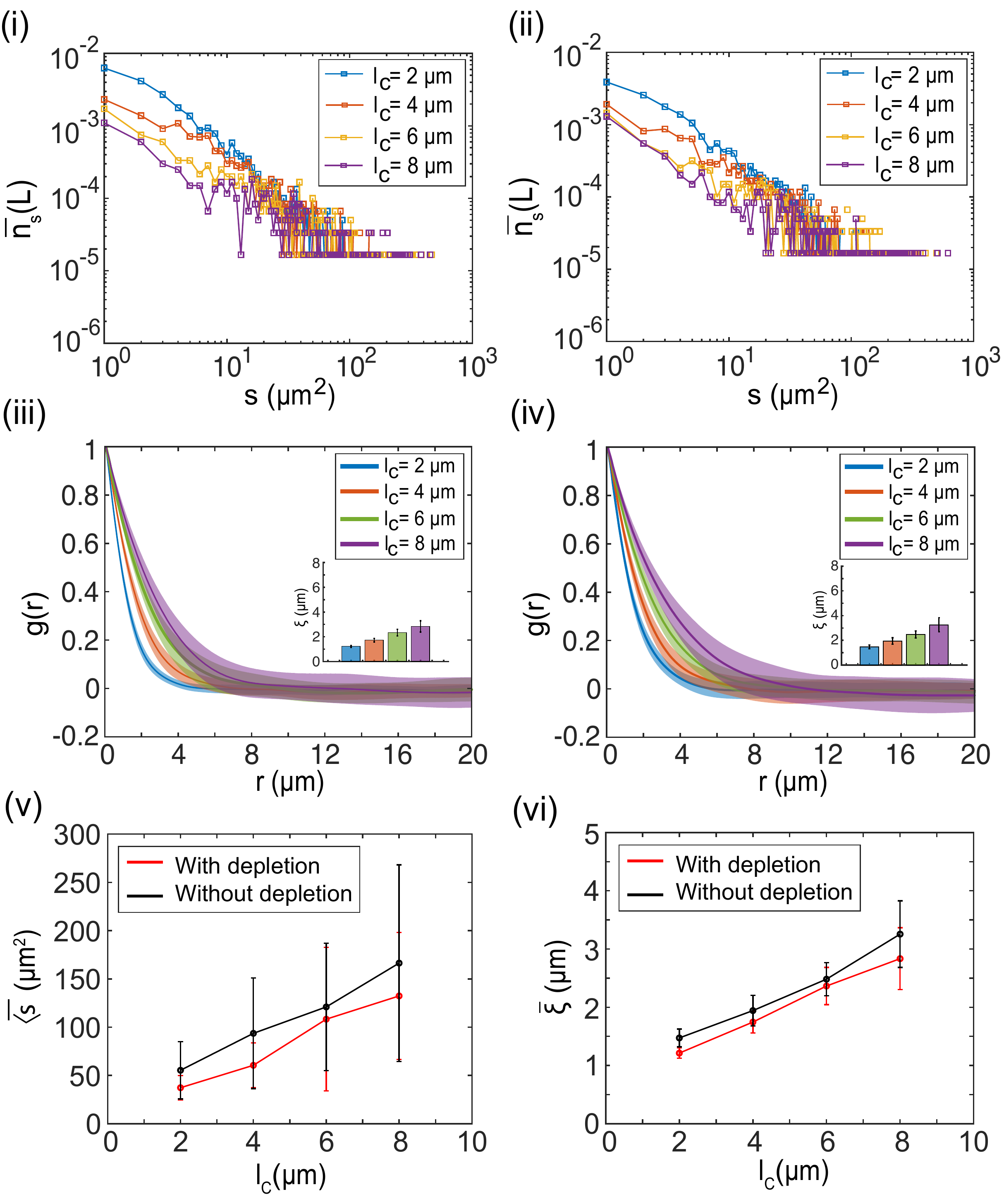}
\caption{\label{fig:dep_no_dep}Data from Cytosim model with attractive force comparing (i) domain size distribution and (iii) radially averaged mean auto correlation function for different contour lengths. Data from Cytosim model without attractive force comparing (ii) domain size distribution and (iv) radially averaged mean auto correlation function comparing different contour lengths. All data shown for $\phi=0.75$ (n=6), $l_c$ is 2 $\mu$m (blue squares), 4 $\mu$m (red squares), 6 $\mu$m (yellow squares), and 8 $\mu$m (purple squares). Comparison of the (v) the average domain size, $\overline{\langle s\rangle }$  and (vi) the mean correlation length, $\overline{\xi}$ for no attractive force (black) and with attractive force (red). }
\end{figure}

In our Cytosim simulation, we implemented an effective depletion force using an attractive spring force, equation (\ref{eq:depletion}). Although this has been used in the literature previously \cite{strubing2020,letort2015geometrical}, a more appropriate model for depletion force is a Derjaguin like approximation. When we repeated our simulations in the absence of the attractive spring force, we find the trend is unaffected. We compare the results from simulations without and with attractive spring force for specific packing fraction to compare over different contour lengths. In Fig.\ref{fig:dep_no_dep} we showed that without attractive depletion force the model still follows the same trend of dependence on the contour length. This is the main take-away, and it is independent of whether we use the attractive force or not. Quantitatively, we find that the average domain size, $\overline{\langle s\rangle }$ and mean correlation length, $\overline{\xi}$ are slightly higher without the attractive force.

\section{Statistical analysis and fit data}\label{appendixB}

%The Kolmogorov-Smirnov test (KS test) is a non-parametric test applied on empirical data to test whether the difference in sample is statistically significant, in another term it tests the hypothesis where two samples came from same distribution or not. We report the p-values after comparing two samples using MATLAB function kstest2. The p-value for these tests are reported below in TABLE(\ref{tab:table5}-\ref{tab:table10}) and TABLE(\ref{tab:table11}-\ref{tab:table16}). These table entries are symmetric with diagonal elements 1.

\begin{table}[!ht]
\caption{\label{tab:table5} P-values from two sample KS test performed on domain area data for different crowders as described in results and discussion (Fig. \ref{fig:fig2}B(i))}
\centering
\begin{ruledtabular}
\begin{tabular}{c c c c c}
\multicolumn{1}{c}
\textrm{}&\textrm{8 kDa}&\textrm{14 kDa}&\textrm{88 kDa}&\textrm{100 kDa}\\
\colrule
8 kDa PEG & & $3.3\times10^{-6}$ & $2.7\times10^{-5}$ & $8.1\times10^{-5}$ \\
14 kDa MC & & & $0.1428$ & $0.6937$\\
88 kDa MC & & & & $0.0749$\\
100 kDa PEG\\
\end{tabular}
\end{ruledtabular}
\end{table}

\begin{table}[!ht]
\caption{\label{tab:table6}P-values from two sample KS test performed on domain area data for different tubulin concentrations as described in results and discussion (Fig. \ref{fig:fig2}B(ii))}
\centering
\begin{ruledtabular}
\begin{tabular}{c c c c}
\multicolumn{1}{c}
\textrm{l} & $1.1~\mu m$ & $2.5~\mu m$ & $6.25~\mu m$\\
\colrule
$1.1~\mu m$ & & $4.0\times10^{-6}$ & $8.2\times10^{-14}$ \\
$2.5~\mu m$ & & & $4.9\times10^{-6}$\\
$6.25~\mu m$ \\
\end{tabular}
\end{ruledtabular}
\end{table}

\begin{table}[!ht]
\caption{\label{tab:table7} P-values from two sample KS test performed on domain area data for Cytosim simulation as described in results and discussion (Fig. \ref{fig:fig2}B(iii))}
\centering
\begin{ruledtabular}
\begin{tabular}{c c c c c}
\multicolumn{1}{c}
\textrm{l} & $2~\mu m$ & $4~\mu m$ & $6~\mu m$ & $8~\mu m$\\
\colrule
$2~\mu m$ & & $2.6\times10^{-18}$ & $1.4\times10^{-19}$ & $1.6\times10^{-20}$ \\
$4~\mu m$ & & & $0.011$ & $5.6\times10^{-6}$\\
$6~\mu m$ & & & & $0.0965$\\
$8~\mu m$\\
\end{tabular}
\end{ruledtabular}
\end{table}

\begin{table}[!ht]
\caption{\label{tab:table8} P-values from two sample KS test performed on correlation length for different crowders as described in results and discussion (Fig. \ref{fig:fig2}C(i) inset)}
\centering
\begin{ruledtabular}
\begin{tabular}{c c c c c}
\multicolumn{1}{c}
\textrm{}&\textrm{8 kDa}&\textrm{14 kDa}&\textrm{88 kDa}&\textrm{100 kDa}\\
\colrule
8 kDa PEG & & $0.2464$ & $0.0060$ & $0.0760$ \\
14 kDa MC & & & $0.0067$ & $0.0167$\\
88 kDa MC & & & & $1.7\times10^{-6}$\\
100 kDa PEG\\
\end{tabular}
\end{ruledtabular}
\end{table}

\begin{table}[!ht]
\caption{\label{tab:table9} P-values from two sample KS test performed on correlation length for different tubulin concentration as described in results and discussion (Fig. \ref{fig:fig2}C(ii) inset)}
\centering
\begin{ruledtabular}
\begin{tabular}{c c c c}
\multicolumn{1}{c}
\textrm{l} & $1.1~\mu m$ & $2.5~\mu m$ & $6.25~\mu m$\\
\colrule
$1.1~\mu m$ & & $1.4\times10^{-4}$ & $8.6\times10^{-11}$ \\
$2.5~\mu m$ & & & $2.5\times10^{-19}$\\
$6.25~\mu m$ \\
\end{tabular}
\end{ruledtabular}
\end{table}

\begin{table}[!ht]
\caption{\label{tab:table10} P-values from two sample KS test performed on correlation length for Cytosim data as described in results and discussion (Fig. \ref{fig:fig2}C(iii) inset)}
\centering
\begin{ruledtabular}
\begin{tabular}{c c c c c}
\multicolumn{1}{c}
\textrm{l} & $2~\mu m$ & $4~\mu m$ & $6~\mu m$ & $8~\mu m$\\
\colrule
$2~\mu m$ & & $7.4\times10^{-11}$ & $8.8\times10^{-12}$ & $8.8\times10^{-12}$ \\
$4~\mu m$ & & & $4.0\times10^{-9}$ & $7.4\times10^{-11}$\\
$6~\mu m$ & & & & $0.0029$\\
$8~\mu m$\\
\end{tabular}
\end{ruledtabular}
\end{table}

\begin{table}[!ht]
\caption{\label{tab:table17} Fit parameters for the number scaling and the area scaling in log-log scale using the equation y = ax+b (Fig. \ref{fig:fig11}(i), (iii))}
\centering
\begin{ruledtabular}
\begin{tabular}{c c c c}
\multicolumn{1}{c}
\textrm{} & \textrm{a}\footnotemark[1] & \textrm{b}\footnotemark[1] & \textrm{R-squared}\\
\colrule
Number scaling & $-1.6\pm 0.1$ & $-3.1\pm0.1$ & $0.9803$ \\
\colrule
Area scaling & $1.5\pm 0.1$ & $3.4\pm0.1$ & $0.9831$ \\
\end{tabular}
\end{ruledtabular}
\footnotetext[1]{All data presented are value $\pm$ 95$\%$ confidence bound}
\end{table}

\begin{table}[!ht]
\caption{\label{tab:table18} Normalized absorbance data fit parameters from turbidity measurement,  using eq. (\ref{eq:boltzman}) (Fig. \ref{fig:fig10}(ii))}
\centering
\begin{ruledtabular}
\begin{tabular}{c c c c c c}
\multicolumn{1}{c}
\textrm{} & \textrm{a}\footnotemark[1] & \textrm{b}\footnotemark[1] & $t_{0} (s)$\footnotemark[1] & $\tau(s)$\footnotemark[1] & \textrm{R-squared}\\
\colrule
Tubulin & $0.994$ & $1.05$ & $1266$ & $284.7$ & $0.9952$ \\
&$\pm0.007$ & $\pm0.03$ & $\pm 24$ & $\pm 22.2$ & \\
\colrule
8 kDa & $0.991$ & $1.10$ & $751.4$ & $193.9$ & $0.9949$ \\
&$\pm0.009$ & $\pm0.04$ & $\pm 20.4$ & $\pm 16.1$ & \\
\colrule
14 kDa & $0.989$ & $1.01$ & $1664$ & $228.1$ & $0.9986$ \\
&$\pm0.008$ & $\pm0.01$ & $\pm 10$ & $\pm 9.8$ & \\
\colrule
88 kDa & $0.984$ & $1.01$ & $1377$ & $252.8$ & $0.9985$ \\
&$\pm0.008$ & $\pm0.01$ & $\pm 12$ & $\pm 10.9$ & \\
\colrule
100 kDa & $0.984$ & $1.01$ & $1093$ & $184.1$ & $0.9973$\\
&$\pm0.008$ & $\pm0.02$ & $\pm 13$ & $\pm 11.3$ & \\
\end{tabular}
\end{ruledtabular}
\footnotetext[1]{All data presented are value $\pm$ 95$\%$ confidence bound}
\end{table}

\begin{table}[!ht]
\caption{\label{tab:table11} P-values from two sample KS test performed on tactoid length for different crowders on polymer brush surface as described in results and discussion (Fig. \ref{fig:fig3}(i))}
\centering
\begin{ruledtabular}
\begin{tabular}{c c c c c}
\multicolumn{1}{c}
\textrm{}&\textrm{8 kDa}&\textrm{14 kDa}&\textrm{88 kDa}&\textrm{100 kDa}\\
\colrule
8 kDa PEG & & $0.04$ & $5.2\times10^{-60}$ & $1.1\times10^{-20}$ \\
14 kDa MC & & & $5.2\times10^{-49}$ & $1.1\times10^{-16}$\\
88 kDa MC & & & & $4.6\times10^{-46}$\\
100 kDa PEG\\
\end{tabular}
\end{ruledtabular}
\end{table}

\begin{table}[!ht]
\caption{\label{tab:table13} P-values from two sample KS test performed on tactoid width for different crowders on polymer brush surface as described in results and discussion (Fig. \ref{fig:fig3}(ii))}
\centering
\begin{ruledtabular}
\begin{tabular}{c c c c c}
\multicolumn{1}{c}
\textrm{}&\textrm{8 kDa}&\textrm{14 kDa}&\textrm{88 kDa}&\textrm{100 kDa}\\
\colrule
8 kDa PEG & & $5.0\times10^{-11}$ & $9.5\times10^{-11}$ & $1.9\times10^{-8}$ \\
14 kDa MC & & & $4.1\times10^{-37}$ & $1.5\times10^{-34}$\\
88 kDa MC & & & & $0.8034$\\
100 kDa PEG\\
\end{tabular}
\end{ruledtabular}
\end{table}

\begin{table}[!ht]
\caption{\label{tab:table15} P-values from two sample KS test performed on tactoid aspect ratio for different crowders on polymer brush surface as described in results and discussion (Fig. \ref{fig:fig3}(iii))}
\centering
\begin{ruledtabular}
\begin{tabular}{c c c c c}
\multicolumn{1}{c}
\textrm{}&\textrm{8 kDa}&\textrm{14 kDa}&\textrm{88 kDa}&\textrm{100 kDa}\\
\colrule
8 kDa PEG & & $1.1\times10^{-12}$ & $7.9\times10^{-36}$ & $0.0055$ \\
14 kDa MC & & & $2.3\times10^{-13}$ & $8.9\times10^{-7}$\\
88 kDa MC & & & & $8.8\times10^{-36}$\\
100 kDa PEG\\
\end{tabular}
\end{ruledtabular}
\end{table}

\begin{table}[!ht]
\caption{\label{tab:table12} P-values from two sample KS test performed on tactoid length for different crowders on lipid surface as described in results and discussion (Fig. \ref{fig:fig4-2}(i))}
\centering
\begin{ruledtabular}
\begin{tabular}{c c c c c}
\multicolumn{1}{c}
\textrm{}&\textrm{}&\textrm{100 kDa PEG}&\textrm{}&\textrm{}\\
\colrule
\textrm{} & 88 kDa MC & $7.4\times10^{-27}$&\textrm{}\\
\end{tabular}
\end{ruledtabular}
\end{table}

\begin{table}[!ht]
\caption{\label{tab:table14} P-values from two sample KS test performed on tactoid width for different crowders on lipid brush surface as described in results and discussion (Fig. \ref{fig:fig4-2}(ii))}
\centering
\begin{ruledtabular}
\begin{tabular}{c c c c c}
\multicolumn{1}{c}
\textrm{}&\textrm{}&\textrm{100 kDa PEG}&\textrm{}&\textrm{}\\
\colrule
\textrm{} & 88 kDa MC & $0.6167$&\textrm{}\\
\end{tabular}
\end{ruledtabular}
\end{table}

\begin{table}[!ht]
\caption{\label{tab:table16} P-values from two sample KS test performed on tactoid aspect ratio for different crowders on lipid surface as described in results and discussion (Fig. \ref{fig:fig4-2}(iii))}
\centering
\begin{ruledtabular}
\begin{tabular}{c c c c c}
\multicolumn{1}{c}
\textrm{}&\textrm{}&\textrm{100 kDa PEG}&\textrm{}&\textrm{}\\
\colrule
\textrm{} & 88 kDa MC & $2.5\times10^{-13}$&\textrm{}\\
\end{tabular}
\end{ruledtabular}
\end{table}

\begin{table}[!ht]
\caption{\label{tab:table20} Length vs aspect ratio fit parameters for all tactoids, using the equation $y = mx$ (Fig. \ref{fig:fig8}(i))}
\centering
\begin{ruledtabular}
\begin{tabular}{c c c}
\multicolumn{1}{c}
\textrm{} & \textrm{m}($1/\mu m$)\footnotemark[1] & \textrm{R-squared}\\
\colrule
Initial & $1.21\pm 0.01$ & $0.2573$ \\
\end{tabular}
\end{ruledtabular}
\footnotetext[1]{All data presented are value $\pm$ 95$\%$ confidence bound}
\end{table}

\begin{table}[!ht]
\caption{\label{tab:table21} $\alpha$ vs R fit parameters for all tactoids, using the equation $y = Ax^{\delta}$ (Fig. \ref{fig:fig8}(iii))}
\centering
\begin{ruledtabular}
\begin{tabular}{c c c c}
\multicolumn{1}{c}
\textrm{} & \textrm{A}\footnotemark[1] & $\delta$ \footnotemark[1] & \textrm{R-squared}\\
\colrule
later & $1.3\pm 0.1$ & $-1.69 \pm 0.04$ &$0.8113$ \\
\end{tabular}
\end{ruledtabular}
\footnotetext[1]{All data presented are value $\pm$ 95$\%$ confidence bound}
\end{table}

\begin{table}[!ht]
\caption{\label{tab:table19} Tactoid growth characteristics fit parameters, using the equation $y = ax + b$ (Fig. \ref{fig:dynamics})}
\centering
\begin{ruledtabular}
\begin{tabular}{c c c c}
\multicolumn{1}{c}
\textrm{Parameter} & \textrm{a}\footnotemark[1] & \textrm{b}\footnotemark[1] & \textrm{R-squared}\\
\colrule
Length $(\mu m)$ & $(10.7\pm 0.2)$ & $4.24\pm0.03$ & $0.9892$ \\
(0 s to 200 s)& $\times 10^{-3}$& &\\
\colrule
Length $(\mu m)$& $(0.39\pm 0.02)$ & $8.16\pm0.04$ & $0.7689$ \\
(2000 s to 3600 s)& $\times 10^{-3}$& &\\
\colrule
Width $(\mu m)$& $(0.048\pm 0.002)$ & $1.029\pm0.004$ & $0.6406$ \\
(200 s to 3600 s)& $\times 10^{-3}$ & &\\
\colrule
Aspect ratio & $(8.8\pm 0.6)$ & $4.41\pm0.07$ & $0.89$ \\
(0 s to 200 s)& $\times 10^{-3}$ & &\\
\colrule
Aspect ratio & $(0.11\pm 0.04)$ & $7.7\pm0.1$ & $0.0310$ \\
(2000 s to 3600 s)& $\times 10^{-3}$ & &\\
\end{tabular}
\end{ruledtabular}
\footnotetext[1]{All data presented are value $\pm$ 95$\%$ confidence bound}
\end{table}

\pagebreak

\section{Tactoid Scales}\label{appendixC}
In this appendix we discuss briefly the theoretical aspect of the tactoid assembly previously described ~\cite{Kaznacheev2002TheCrystals,Prinsen2003ShapeTactoids,Kaznacheev2003TheCrystals,Prinsen2004ContinuousTactoids}. Total free energy functional of a nematic tactoid can be written in general form as, 
\begin{equation}
F=F_{E}+F_{S}  
\end{equation}

where $F_{E}$ is the bulk elastic energy, accounts for director field deformation inside the tactoid, and $F_{S}$ is the interfacial energy term. The elastic term in the Frank energy, which is integrated over volume $V$ can be written down as,  
\begin{multline*}
F_{E} = \int_{V} d^{3}{\bf r}\biggl[ \frac{K_{1}}{2}({ \bf \nabla} \cdot {\bf n})^{2}+ \frac{K_{2}}{2}({\bf n}\cdot{\bf \nabla} \times {\bf n})^{2} \\
+\frac{K_{3}}{2}({\bf n}\times{\bf \nabla} \times {\bf n})^{2} - K_{24}{\bf \nabla}\cdot[ {\bf n}{\bf \nabla}\cdot{\bf n} + {\bf n}\times({\bf \nabla}\times{\bf n})] \biggr]  
\end{multline*}
in terms of director field ${\bf n(r)}$. Here, $K_{1}$, $K_{2}$, $K_{3}$, $K_{24}$ are splay, twist, bend and saddle-splay deformation mode, respectively. Usually, the twist term $K_{2}$ is dropped assuming a no-twist condition. The $K_{24}$ term can be absorbed into $K_{1}$ to give:
\begin{equation}\label{A1}
F_{E} = \int_{V} d^{3}{\bf r}\left[\frac{K_{1}}{2} ({ \bf \nabla} \cdot {\bf n})^{2} + \frac{K_{3}}{2}({\bf n}\cdot{\bf \nabla} \times {\bf n})^{2} \right]
\end{equation}
The interfacial surface energy term, which is integrated over the tactoid surface, has two components with accompanying parameters. The first component comes from the isotropic surface energy, which is driven by the interfacial tension, $\tau$. The second term introduces anisotropy in the surface energy term, which is driven by the ratio of the anchoring strength and interfacial tension, $\omega$, a measure how directors interact with the interface.
\begin{equation}\label{A2}
F_{S} = \tau \int_{S} d^{2}{\bf r}\{ 1 + \omega({\bf q}\cdot{\bf n})^{2}\} 
\end{equation}
Here ${\bf q}$ is the unit normal to the interfacial surface.
As described, this system can be parametrized using $R$ and $\alpha$. From a simple scaling argument, it can be shown that, the bulk elastic deformation cost should scale as $\sim KR^{3}$ multiplied by $\frac{1}{R^{2}}$ which accounts for square derivatives, i.e. $\sim KR$ and surface anchoring energy will have $\sim \omega R^{2}$ scaling. The total free energy, written in terms of these two independent variables, takes the form:
\begin{eqnarray*}
F&=&F_{E}+F_{S}  \\
&=& \sum_{i=1,3} K_{i}Rf_{E}^{(i)}(\alpha)+\tau R^{2}(f_{S}(\alpha)+\omega f_{W}(\alpha))
\end{eqnarray*}
\begin{equation}
\implies \widetilde{F} = \sum_{i=1,3}\kappa_{i} \psi^{(i)}_{E}(\alpha) + \psi_{S}(\alpha) + \omega \psi_{W}(\alpha)
\end{equation}
where $\widetilde{F}=\frac{F}{\tau V^{\frac{2}{3}}}$, $\kappa_{i} = \frac{K_{i}}{\tau V^{\frac{1}{3}}}$, $\psi_{E}^{(i)}(\alpha)=\frac{f_{E}^{(i)}(\alpha)}{v(\alpha)^{\frac{1}{3}}}$, $\psi_{W,S}(\alpha)=\frac{f_{W,S}(\alpha)}{v(\alpha)^{\frac{2}{3}}}$ are the scaled variables. The surface area and volume are defined as $A=R^{2}f_{S}(\alpha)$ and $V=R^{3}v(\alpha)$
\begin{eqnarray*}
f_{S}(\alpha) &=& 4\pi (\sin\alpha - \alpha \cos\alpha)\\
{\it v}(\alpha) &=& 2\pi (\sin\alpha - \alpha \cos\alpha - \frac{\sin^{3}\alpha}{3})
\end{eqnarray*}
The director field inside a tactoid is related to its size or volume. There is a cross-over from the homogeneous to the bipolar director field as volume gets larger, predicted from theory ~\cite{Prinsen2003ShapeTactoids} and verified in the experiments with carbon nanotubes in chlorosulfonic acid ~\cite{Jamali2015ExperimentalTactoids}.  The homogeneous director field configuration is elastically rigid and has an unperturbed director, ${\bf n(r)}={\bf n}$. In the small volume limit, $V\to 0$, $1\ll \omega < \kappa$, therefore the elastic deformation term, $\psi_{E}^{(i)}(\alpha)$ must be zero as it introduces a huge contribution from a small perturbation in the director field. Only surface terms remain with $\omega$ being the only scaled parameter in the problem. This condition defines the crossover volume, $V\sim\left( \frac{K}{\tau\omega}\right)^{3}$. From the Wulff construction, it has been predicted that aspect ratio $l/r=2\omega^{1/2}$ for $\omega \gg 1$.\\

\bibliographystyle{unsrt}
\bibliography{references1.bib,references2.bib}

\end{document}